\documentclass[aps,10pt,tightenlines,twocolumn,twoside,showpacs,nofootinbib,superscriptaddress]{revtex4-2}
\usepackage{amsfonts}
\usepackage{float}
\usepackage{placeins}
\usepackage{graphicx}
\usepackage[hidelinks]{hyperref}
\usepackage{color,amsmath,amssymb,bm}
\usepackage{tensor}
\usepackage[normalem]{ulem}  
\usepackage{changepage}   
\usepackage{slashed}
\usepackage{siunitx}
\usepackage{mathbbol}
\usepackage{bbm}
\usepackage{hyperref}
\DeclareSymbolFontAlphabet{\mathbbmm}{bbold}
\DeclareMathAlphabet{\mathbbmsl}{U}{bbm}{m}{sl}
\DeclareSymbolFontAlphabet{\mathbb}{AMSb}%

\usepackage[normalem]{ulem}

\def \no {\nonumber}
\def \a {\alpha}
\def \b {\beta}

\def \d {\delta}
\def \e {\epsilon}
\def \g {\gamma}

\NewCommandCopy{\oldo}{\o}
\DeclareRobustCommand{\o}{\ifmmode\omega\else\oldo\fi}

\def \r {\rho}
\def \l {\lambda}
\def \m {\mu}
\def \n {\nu}
\def \s {\sigma}

\def \pd {\partial}

\def \<{\langle}
\def \>{\rangle}
\def \+{\dagger}

\def \[{\left[}
\def \]{\right]}

\def \vx {\bm{x}}

\def \vk {\bm{k}}

\def \ve {\varepsilon}
\def \<{\langle}
\def \>{\rangle}
\def \+{\dagger}

\def \[{\left[}
\def \]{\right]}

\def \srho 
{{\tilde \rho}}

\renewcommand{\sout}{\bgroup \color{red} \ULdepth=-.5ex \ULset}

\def\blfootnote{\xdef\@thefnmark{}\@footnotetext}

\newcommand{\beq}{\begin{equation}}
\newcommand{\eeq}{\end{equation}}
\newcommand{\bea}{\begin{eqnarray}}
\newcommand{\eea}{\end{eqnarray}}

\begin{document}
\title{
	Deriving the anomalous Lorentz transformation and side jump using a trace method
}
\author{Jin-Rui Wu}
\affiliation{School of Physics\&Electronics, Hunan University, Changsha 410082, China}
\author{Feng Li}
\affiliation{School of Physics\&Electronics, Hunan University, Changsha 410082, China}
\affiliation{Hunan Provincial Key Laboratory of High-Energy Scale Physics
	and Applications, Hunan University, Changsha 410082, China}
\author{Shuai Y.F.~Liu}
\email{lshphy@hnu.edu.cn}
\affiliation{School of Physics\&Electronics, Hunan University, Changsha 410082, China}
\affiliation{Hunan Provincial Key Laboratory of High-Energy Scale Physics
	and Applications, Hunan University, Changsha 410082, China}

\date{\today}

\begin{abstract}
The side jump in the anomalous Lorentz transformation, arising from the spin-orbit interactions, plays important roles in various intriguing physics, such as chiral vortical effects and spin polarization. In this work, the side jump of the spin-half particles is visualized and evaluated ab initio within Dirac theory, through a comprehensive study of the density $\psi^\dagger\psi$ and position $\int d^3x\psi^\dagger \mathbf{x}\psi$ of a wave packet under Lorentz transformation.  A compact gamma-matrix representation of the Lorentz boost matrix of a Dirac spinor, $S(\bm{\eta})= h_{\bm\eta}^{\mu}\gamma_\mu \check{u}^\nu \gamma_\nu$, has been found.
Based on this expression, a trace method has been developed to derive the anomalous Lorentz transformation of the center of the probability density of a spin-half fermion wave packet, and a compact analytical expression for the side jump has been obtained through a transparent and tractable derivation. The momentum width plays an important role in the side jump. For example, for Gaussian wave packets, increasing the width suppresses the displacement, while the side jump at fixed width depends nonmonotonically on mass, reaching a maximum at a finite mass.  In the zero-width limit, the side jump takes a compact form perpendicular to both the Pauli–Lubanski vector and the boost direction, recovering the known result for a chiral particle.



\end{abstract}

\maketitle
\section{Introduction}
The side jump, i.e., an anomalous displacement arising from the spin-orbit interactions, was first introduced for dealing with the scattering processes in the anomalous Hall effect~\cite{PhysRevB.2.4559,RevModPhys.82.1539,RevModPhys.87.1213}. Given the relativistic nature of the spin-orbit interactions, it is demonstrated in Refs.~\cite{Chen:2014cla,Duval:2014cfa,Stone:2015kla,Chen:2015gta,Hidaka:2016yjf} that the side jumps in the scattering processes are related to the anomalous Lorentz transformation, where the space-time coordinates of the chiral particles transform as $x^{\prime\mu}=\Lambda^\mu_{~\nu}x^\nu+\Delta^\mu$, with the spin-dependent displacement $\Delta^\mu$ being essential for preserving the covariance of the total angular momentum ~\cite{Chen:2014cla,Chen:2015gta}. 
As a shift in an orbital quantity, i.e., the particle position $x$, $\Delta^\mu$ connects the orbital angular momentum with spin, and is therefore the bridge via which the orbital and the spin angular momenta communicate under a Lorentz transformation. Consequently, the side jumps play a key role in the microscopic explanation of the chiral vortical effect~\cite{Vilenkin:1979ui,Son:2009tf,Kharzeev:2010gr,Kharzeev:2015znc} and the macroscopic polarization occurring in a vortical fluid due to the spin-orbit interaction.
Such theoretical insights have  been implemented, for the first time, in the transport model~\cite{Liu:2019krs}, which respects the angular momentum conservation law, for simulating the heavy-ion collisions. As expected, the chiral vortical effect is dynamically generated in the simulation. This approach also provides a promising microscopic  perspective for understanding the spin puzzle in the heavy-ion collisions, i.e., the mismatch between the theoretical prediction~\cite{Becattini:2017gcx} and the experimental measurements~\cite{Niida:2018hfw,Adam:2019srw} of the local $\Lambda$ polarization.

Although the side jumps and related physics have been extensively studied using a wide variety of methods in different  contexts~\cite{born1935quantization,pryce1948mass,moller1949definition,Newton:1949cq,PhysRev.137.B188,Chen:2014cla,Duval:2014cfa,Stone:2015kla,Chen:2015gta,Hidaka:2016yjf,bliokh2017position,Lorce:2018zpf,Hattori:2019ahi}, the final results are often obtained under specific assumptions or within particular formulations, such as taking the massless limit or redefining the position operators.
In addition, these methods  usually involve specialized knowledge and advanced techniques, which are not always accessible to general readers and limit the applicability of these methods to phenomenological studies. 
On the other hand, the quantity $ \int d^3 x\,\psi^\dagger \mathbf{x} \psi $, a textbook expression for the position expectation value in Dirac theory, seems not to have been thoroughly studied especially for its properties under Lorentz transformation, where essential physics and subtleties of the side jump can be mostly illustrated through this straightforward evaluation.  
Meanwhile, as the Lorentz transformation of $ \int d^3 x\,\psi^\dagger \mathbf{x} \psi $ is not manifest, the straightforward evaluation becomes complicated and a more tractable method for evaluating this kind of non-covariant quantity should be useful.

With these motivations in mind, we have found a compact gamma-matrix representation for the Lorentz boost $S(\boldsymbol\eta)= h_{\bm\eta}^{\mu}\gamma_\mu \check{u}^\nu \gamma_\nu\;$. With the help of this expression, we developed a  trace method to evaluate the Lorentz transformation of $ \int d^3 x\psi^\dagger \mathbf{x} \psi$ ab initio, with all steps elaborated  in an analytical and tractable way.  
A compact expression for the anomalous Lorentz transformation, including side-jump effects, is obtained exactly within the free Dirac wave-packet framework, without a semiclassical expansion.

\section{Derivation from Dirac Waves}
\label{sec_wavecal}
\subsection{Wave packets and their probability centers}
\begin{figure}[!h]
	\includegraphics[width=0.5\textwidth]{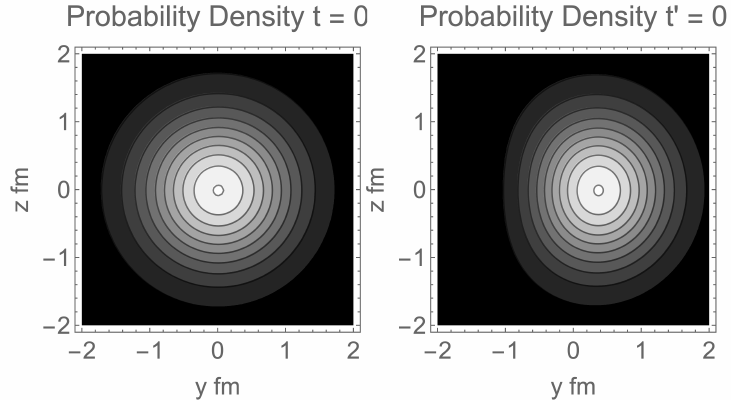}
	\caption{Probability density at $x=x'=0$ of a massive spinor polarized in the positive z-direction in the particle rest frame (left) and the frame boosted in the positive x-direction (right)}
	\label{fig:visualization}
\end{figure}

Before  proceeding to the derivation, we begin with an intuitive visualization. We plot the probability density, defined as $\psi^\dagger\psi$, of a spinor wave-packet polarized in the positive z-direction in  the particle rest frame and in the frame boosted in the positive x-direction with a rapidity equal to 4 in the left and right panels of Fig.~\ref{fig:visualization}, respectively. 

To characterize the phenomena observed in Fig.~\ref{fig:visualization} with more concrete formulas, we will first express the spinor wave packet in its conventional form~\cite{Peskin:1995ev},
\begin{equation}
\label{eq:wavepackage}
    \psi(x) = \int \frac{d^3 \mathbf k}{(2\pi)^3 \sqrt{2 \varepsilon_k}} a(k) u(k) e^{-i(\varepsilon_k t - \mathbf k \cdot \mathbf x)}\;,
\end{equation}
where
\begin{equation}
\label{eq:DiracSpinor}
    u(k)=\sqrt{m} S(-\boldsymbol\eta_k)\left(
    \begin{tabular}{c}
         $\xi$  \\
         $\xi$ 
    \end{tabular}
    \right)
\end{equation}
is the solution to the Dirac equation in the momentum representation with $\boldsymbol\eta_k$ being related to $\mathbf k$ by $\mathbf k = m\sinh|\eta_k|\hat{\boldsymbol\eta}_k$.  With $ \check{u}=(1,\bm0),h_{\bm\eta}=\Lambda_{\boldsymbol\eta/2} \check{u},u_{\bm\eta}=\Lambda_{\boldsymbol\eta} \check{u}$, a useful gamma-matrix representation of the transformation matrix has been found:
\begin{equation}
	\label{eq:sgamma}
    S(\boldsymbol\eta)= h_{\bm\eta}^{\mu}\gamma_\mu \check{u}^\nu \gamma_\nu,\; h=(u_{\bm \eta}+\check{ u})/\sqrt{2(u_{\bm\eta}\cdot \check{u}+1)}.
\end{equation}
$S(\boldsymbol\eta)$ is the transformation matrix which is not only used in Eq. (\ref{eq:DiracSpinor}) but also employed for transforming a spinor wave-packet as
\begin{align}
    \psi^\prime(x)=S(\boldsymbol\eta)\psi\left(\Lambda^{-1}_{\boldsymbol\eta} x\right) 
    \label{eq:psiprime}
\end{align} 
under a boost with rapidity $\boldsymbol\eta$, where
\begin{equation}
	\Lambda_{\boldsymbol{\eta}} = \left(
	\begin{array}{cc}
		\cosh \eta & -\sinh \eta \hat{\boldsymbol{\eta}} \\
		-\sinh \eta \hat{\boldsymbol{\eta}} & I + (\cosh \eta - 1)\hat{\boldsymbol{\eta}}\hat{\boldsymbol{\eta}}
	\end{array}
	\right).
\end{equation}
 The polarization direction of the wave-packet in its rest frame is determined by $\xi$, which satisfies $\xi^\dagger\boldsymbol{\sigma}\xi=\hat{\mathbf n}$ with $\hat{\mathbf n}$ being the spin direction of the fermionic wave-packet in its rest frame, and $\boldsymbol\sigma$ being the Pauli matrices. In the general expressions below, $\xi$ and $\hat{\mathbf n}$ may depend on $\mathbf k$, and $u(0)$ denotes the rest spinor constructed from the corresponding $\xi$.
 $\hat{\mathbf n}$ is chosen along the z-axis when we plot Fig. \ref{fig:visualization}. $a(k)=a_0(k)e^{i\phi(k)}$ is the c-number amplitude for a given momentum $\mathbf k$ and satisfies the normalization condition that
 \begin{equation}
     \int d^3\mathbf x \psi^\dagger \psi = \int \frac{d^3 \mathbf k}{(2\pi)^3} |a(k)|^2 =1,
 \end{equation}
 and is chosen to be Gaussian, i.e., $\mathcal{N} e^{-\frac{|\mathbf k|^2}{2\sigma^2}}$, for plotting Fig. \ref{fig:visualization}, where the mass and the smearing width $\sigma$ are taken as $0.5$ and $1$ fm$^{-1}$, respectively. All the above and the following calculations are carried out in the Weyl representation, while the physics should be independent of the choice of  representation. As shown in Fig. \ref{fig:visualization}, the wave-packet in the boosted frame is shifted in the positive y-direction, which is perpendicular to both the spin direction in the particle rest frame and the direction of the boost, indicating the existence of a side jump $\boldsymbol\Delta \propto \hat{\mathbf n} \times \boldsymbol \eta$. The side jump of a massive fermion is thus confirmed and visualized, and will be evaluated as follows.

Under the assumption that $\lim_{|\mathbf k|\to\infty}a(k) = 0$, the expectation value of the spatial coordinate of a boosted spinor wave-packet, evaluated using the $\psi^\prime(x)$ defined in Eq. (\ref{eq:psiprime}), is
\begin{eqnarray}
\label{eq:x1}
    \bar{\mathbf x}_{\boldsymbol \eta}(t) &=& \int d^3\mathbf x\; \mathbf x\, \psi^{\prime\dagger}(x)\psi^\prime( x) \nonumber\\
    &=& i \int \frac{d^3\mathbf k^\prime}{(2\pi)^3} \widetilde\psi^{\prime\dagger}(\mathbf k^\prime,t) \partial_{\mathbf k^\prime}\widetilde\psi^\prime(\mathbf k^\prime,t)
\end{eqnarray}
where
\begin{eqnarray}
	\label{eq:psikp}
    \widetilde\psi^\prime(\mathbf k^\prime, t) &\equiv& \int d^3\mathbf x \psi^\prime(x) e^{-i\mathbf k^\prime \cdot \mathbf x} \nonumber\\  
    &=& S(\boldsymbol{\eta}) \frac{\varepsilon_k}{\varepsilon_{k^\prime}} \frac{a(k)}{\sqrt{2\varepsilon_k}} u(k) e^{-i\varepsilon_{k^\prime} t}.
\end{eqnarray}
Here, $k$ is related to $k^\prime$ by a Lorentz transformation, namely, $k^\prime = \Lambda_{\boldsymbol\eta} k$ and $\mathbf{k}'$ is the spatial part of $k'$. Note that the expression in Eq.~(\ref{eq:psikp}) contains both $k$ and $k'$. In addition, the identity
\begin{align}
\label{eq:ussu}
u^{\dagger}_s(k)S^2(\eta) u_{s'}(k)=2\varepsilon_{k'}\delta_{ss'}
\end{align}
 is a useful  for the subsequent derivation, where $s$ and $s'$ is the index for spin basis.
Under the change of integration variables from $\mathbf{k}^\prime$ to $\mathbf{k}$, the integration measure $d^3 \mathbf k^\prime / (2 \varepsilon_{k^\prime}) = d^3 \mathbf k / (2 \varepsilon_k)$ remains invariant, and Eq. (\ref{eq:x1}) can thus be written after some efforts:
\begin{eqnarray}
\label{eq:x2}
    \bar{\mathbf x}_{\boldsymbol \eta}(t)
    &=& \bar{\mathbf v}_{\boldsymbol{\eta}}t +i\int \frac{d^3\mathbf k}{(2\pi)^3} \Big[
    \frac{a^\ast(k)}{\sqrt{2\varepsilon_k}} u^\dagger(0)S(-\bm{\eta}_{k}) \nonumber\\
    &&S^2(\boldsymbol\eta) \partial_{\mathbf k^\prime}\left(\frac{\varepsilon_k}{\varepsilon_{k^\prime}}\frac{a(k)}{\sqrt{2\varepsilon_k}} S(-\bm{\eta}_{k})u(0)\right)\Big]\nonumber\\
    &=& \mathbf d_{\bm \eta}+ \bar{\mathbf v}_{\boldsymbol{\eta}}t+\delta\mathbf x_{\boldsymbol\eta},
\end{eqnarray}
where
\begin{equation}
	\label{eq:deta}
	\mathbf{d}_{\bm \eta} \equiv \int \frac{d^3\mathbf k}{(2\pi)^3}|a(k)|^2[-\partial_{\mathbf k^\prime}\phi+i\xi^\dagger\partial_{\mathbf k'}\xi]
\end{equation}
is a gauge-invariant contribution from the phase gradient and the Berry connection $i\xi^\dagger\partial_{\mathbf k'}\xi$.
We need the identity in Eq.~(\ref{eq:ussu}) to extract the inner product $\xi^\dagger\pd_{\mathbf{k}'}\xi$ from the expression containing $S^2$ between the spinors.
The velocity of the charge center of the wave packet, $\bar{\mathbf v}_{\boldsymbol\eta}$, is

\begin{equation}
    \bar{\mathbf v}_{\boldsymbol{\eta}} \equiv \int \frac{d^3\mathbf k}{(2\pi)^3} |a(k)|^2 \frac{\mathbf k^\prime}{\varepsilon_{k^\prime}}
\end{equation}

For $\delta \mathbf x_{\boldsymbol\eta}$, the calculation and simplification are usually non-trivial and cumbersome. In our first online version, we developed a Mathematica code to automate the simplification, reducing a complicated initial expression with a length of several pages to a single line. However, we believe that a simplification method that makes the analytical reduction explicit and can be carried out by hand would still be of interest. Therefore, we developed a trace method, as presented below.
\subsection{Trace evaluation of the anomalous displacement}
Within this trace method, the imaginary contributions will eventually cancel each other, which will be detailed in App.~\ref{sec_details}, while the part containing a nonzero real contribution to $\delta\mathbf{ x}_{\bm \eta}$ in Eq.~(\ref{eq:x2}) can be expressed as: 
\begin{align}
	\label{eq:deltax}
	&\delta\mathbf{ x}_{\bm \eta}
	= \text{Re} \Big [i\int \frac{d^3\mathbf k}{(2\pi)^3} \frac{|a(k)|^2}{2\varepsilon_k'}X(k) \Big ]\\
	\label{eq:tracetech}
	&X(k)=\text{Tr}\left\{
	u(0)u^\dagger(0)\slashed{h}_{-\bm\eta_{k}}\slashed{\check{u}}\slashed{h}_{\bm\eta}\slashed{\check{u}}\slashed{h}_{\bm\eta} \slashed{\check{u}}\partial_{\mathbf k^\prime}\slashed{h}_{-\bm\eta_{k}}\slashed{\check{u}}\right\}\nonumber\\
	&\qquad\;=\frac{m}{2}\text{Tr}\left\{
	(1+\slashed{\check{u}}\gamma_5\slashed{n})\slashed{h}_{\bm\eta_{k}}\slashed{h}_{\bm\eta}\slashed{h}_{-\bm\eta} \partial_{\mathbf k^\prime}\slashed{h}_{-\bm\eta_{k}}\right\},
\end{align}
 where the transformation matrices are represented by gamma matrices using Eq.~(\ref{eq:sgamma}). 
 We also need to note that  $u(0)u^\dagger(0)=m(1+\slashed{\check{u}})(1+\slashed{\check{u}}\gamma_5\slashed{n})/2 $ with $n^\mu\equiv(0, \hat{\mathbf{n} })$, and then
 where we have used the fact that $\slashed{h}_{-\bm\eta_x}\slashed{\check{u}}=\slashed{\check{u}}\slashed{h}_{\bm\eta_x}$. The contribution of ``1" term to the $\delta\mathbf{ x}_{\bm \eta}$  is imaginary and will cancel with other imaginary contributions in the end (see App.~\ref{sec_details} for details). Only the ``$\gamma_5$" term $X_2(k)$  contributes to the final result of $\delta\mathbf{ x}_{\bm \eta}$ and can be straightforwardly expressed as
\begin{align}\label{second_item}
	&\delta\mathbf{ x}_{\bm \eta}=i\int \frac{d^3\mathbf k}{(2\pi)^3} \frac{|a(k)|^2}{2\varepsilon_k'}X_2(k)\no\\
	&X_2(k)=\frac{m}{2}\text{Tr}(\gamma^\nu\gamma^\rho\gamma^\sigma
	\gamma^\alpha\gamma^\beta\gamma^\mu\gamma^5)\no\\
	&\qquad\qquad\times{\check{u}}_\mu n_\nu h_{\bm\eta_k,\rho}h_{\bm\eta,\sigma}h_{-\bm\eta,\alpha}\partial_{\mathbf{k}^\prime} h_{-\bm\eta_k,\beta}.
\end{align}
Hence, the problem reduces to calculating the trace of the product of six gamma matrices with $\gamma^5$. After a basic trace calculation, we obtain
\begin{align}
	\label{eq_trace}
	&\text{Tr}\left(\gamma^\nu\gamma^\rho\gamma^\sigma\gamma^\alpha\gamma^\beta\gamma^\mu\gamma^5\right)=4i\left(\eta^{\alpha\beta}\epsilon^{\mu\nu\rho\sigma}+\eta^{\mu\beta}\epsilon^{\alpha\nu\rho\sigma}\right.\nonumber\\
	&\left.-\eta^{\alpha\mu}\epsilon^{\beta\nu\rho\sigma}-\eta^{\nu\rho}\epsilon^{\sigma\alpha\beta\mu}-\eta^{\sigma\rho}\epsilon^{\nu\alpha\beta\mu}+\eta^{\nu\sigma}\epsilon^{\rho\alpha\beta\mu}\right).
\end{align}
The above calculation uses the identity $\g^{\a}\g^{\b}\g^{\m}=\eta^{\a\b}\g^{\m}+\eta^{\b\m}\g^{\a}-\eta^{\a\m}\g^{\b}-i\e^{\l\a\b\m}\g_{\l}\g^5$, together with the standard trace formula. 
The Schouten identity  $\eta^{\alpha[\beta}\epsilon^{\mu\nu\rho\sigma]}=0$ guarantees that apparently different expressions obtained using cyclic permutations of gamma matrices in the trace operation do not alter the final result. Inserting Eq.~(\ref{eq_trace}) into Eq.~(\ref{second_item}), we get
\begin{align}\label{eq:eta epsilon}
	X_2(k)=
	&i(2m)\left(\eta^{\alpha\beta}\epsilon^{\mu\nu\rho\sigma}+\eta^{\mu\beta}\epsilon^{\alpha\nu\rho\sigma}-\eta^{\alpha\mu}\epsilon^{\beta\nu\rho\sigma}\right.\nonumber\\
	&~\left.	-\eta^{\nu\rho}\epsilon^{\sigma\alpha\beta\mu}-\eta^{\sigma\rho}\epsilon^{\nu\alpha\beta\mu}+\eta^{\nu\sigma}\epsilon^{\rho\alpha\beta\mu}\right)\no\\
	&\times{\check{u}}_\mu n_\nu h_{\bm\eta_k,\rho}h_{\bm\eta,\sigma}
	h_{-\bm\eta,\alpha}\partial_{\mathbf{k}^\prime} h_{-\bm\eta_k,\beta}.
\end{align} 
For the $h_{\mp\bm\eta_{k}}$, there is a simple expression 
\begin{align}
	h_{\mp\bm\eta_{k}}^\mu
	=\left(\pm \bar k^\mu+(m+\varepsilon_{ k})\check{u}^\mu\right)/\sqrt{2(\varepsilon_{ k}+m)m},
\end{align}
where $\bar{k}^\mu=\bar{\d}^{\mu\nu}k_\nu\equiv(0,\bm{k})$ and the projector is defined as $\bar{\delta}^{\mu\nu}=\eta^{\mu\nu}-\check{u}^\mu\check{u}^\nu$ . Therefore, the derivatives can be expressed as
\begin{align}
	&\frac{\partial h^\mu_{-\bm\eta_{k}}}{\partial\bar{k}'^i}
	=\frac{\partial \bar{k}^j}{\partial \bar{k}^{\prime i}}\frac{\partial}{\partial\bar{k}^j}\frac{\bar{k}^\mu+(m+\varepsilon_{ k})\check{ u}^{\mu}}{\sqrt{2(\varepsilon_{ k}+m)m}}.
	\no\\
	&=M^{j}_{\;\;i}\left[ \left( \bar{\delta}^{\mu}_{ j}+\dfrac{\bar{k}^\mu \bar{k}_{j}}{2\varepsilon_k(\varepsilon_k+m)}\right) -\frac{\check{u}^{\mu}\bar{k}_{j}}{2\varepsilon_{k}} \right] D^{-1}_{h_k}
\end{align}
where $M^{j}_{\;\;i}=(\partial \bar k^j)/(\partial \bar k^{\prime i})=[\Lambda_{-\bm\eta}]^{j}_{\;i} -[\Lambda_{-\bm\eta}]^j_{\;0}\bar{k}'_{ i}/\varepsilon_{k'}$ and $D_{h_k}=\sqrt{2m(\ve_k+m)}$.
The last derivative can be obtained straightforwardly. Also, the trace guarantees that the results will be transformed properly under rotations but 
the explicit appearance of the fixed reference vector $\check{u}$ makes the frame dependence explicit.

Notably, Eq. (\ref{eq:eta epsilon}) indicates that the expression for $\delta \mathbf x_{\boldsymbol\eta}$ can be formulated using vector dot and cross products. After a series of simplifications that are detailed in App.~\ref{sec_details}, the resulting expression becomes
\begin{align}
    \label{eq:dx}
    &\delta \mathbf x_{\boldsymbol\eta} \equiv \int \left. \frac{d^3\mathbf k}{(2\pi)^3 2\varepsilon_{k^\prime}}
    |a(k)|^2 \left[\sinh \eta (\hat{\mathbf n}\times\hat{\boldsymbol\eta})\right.\right.\nonumber\\
    &\qquad\left.-\frac{\cosh \eta}{\varepsilon_k+m}(\hat{\mathbf n}\times\mathbf k)+\frac{\cosh \eta-1}{\varepsilon_k+m}(\hat{\mathbf n}\times\mathbf k)\cdot\hat{\boldsymbol\eta} \hat{\boldsymbol\eta}\right]\\
    \label{ed:xbars}
   &\bar{\mathbf{x}}_{\bm \eta}(t)=\mathbf d_{\bm \eta}+ \bar{\mathbf v}_{\boldsymbol{\eta}}t+\delta\mathbf x_{\boldsymbol\eta}
\end{align}
The $\delta \mathbf x_{\boldsymbol\eta} $ is a time-independent anomalous displacement which is related to the spin direction. It needs to be clarified that $\delta \mathbf x_{\boldsymbol\eta}$ is not the side jump we are looking for, since $\delta \mathbf x_{\boldsymbol\eta=\mathbf 0}$ is generally nonzero, while the side jump vanishes by definition at $\boldsymbol \eta = \mathbf 0$ corresponding to the case where no boost is carried out at all. The position of the wave-packet center in the boosted frame is re-expressed compactly in Eq.~(\ref{ed:xbars}). The side jump is obtained by subtracting the corresponding position on the Lorentz-transformed original center trajectory, with both positions evaluated on the chosen equal-time hypersurface. This will be detailed in the next part.

\begin{figure}[!htb]
	\includegraphics[width=0.32\textwidth]{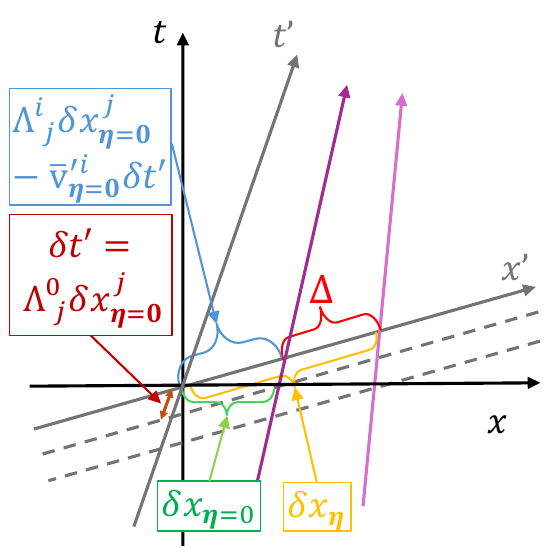}
	\caption{
		Schematic illustration of Eqs.~(\ref{eq:transform}) and (\ref{eq:jump}) . Black and gray axes denote the original and boosted frames. The dark purple line shows the Lorentz-transformed original probability-center trajectory; the light purple line shows the  center trajectory evaluated in the boosted frame. Their spatial separation on $t'=0$ defines the side jump ($\bm\Delta$). For finite-width wave packets, the two trajectories need not be parallel. Physics is similar if we include the $\mathbf{d}$. 
	}
	\label{fig:lineshift}
\end{figure}
\subsection{Definition on an equal-time hypersurface}
Due to the appearance of $\delta \mathbf x_{\boldsymbol\eta}$, the world-line in the boosted frame $(t,\bar{\mathbf x}_{\boldsymbol\eta}(t))$ does not overlap with the one in the original frame $(t,\bar{\mathbf x}_{\boldsymbol\eta=\mathbf 0}(t))$. Thus, we need a Poincar\'e-like transformation rather than homogeneous Lorentz transformation to relate the points on these two world lines.
\begin{equation}
\label{eq:transform}
    \left(
    \begin{tabular}{c}
         $t^\prime$  \\
         $\bar{\mathbf x}_{\boldsymbol\eta}(t^\prime)$
    \end{tabular}
    \right)
    =\Lambda_{\boldsymbol\eta}
    \left(
    \begin{tabular}{c}
         $t_{*}$  \\
         $\bar{\mathbf x}_{\boldsymbol\eta=\mathbf 0}(t_{*})$
    \end{tabular}
    \right)+
    \left(\begin{tabular}{c}
    	$0$  \\
    	$\boldsymbol{\Delta}$
    \end{tabular}\right),
\end{equation}
Here, $t_*$ is chosen such that the Lorentz-transformed point on the original center trajectory has time coordinate $t'$, i.e., $[\Lambda_{\boldsymbol\eta}(t_*,\bar{\mathbf x}_{\boldsymbol\eta=\mathbf 0}(t_*))]^0=t'$. The two events therefore lie on the same equal-time hypersurface, and their separation is purely spatial in the boosted frame. This additional displacement $\boldsymbol{\Delta}$ is the so-called side jump. This choice of a purely spatial separation is consistent with the gauge employed in Refs.~\cite{Chen:2014cla,Chen:2015gta}.
For finite-width wave packets, the side jump can depend on the choice of equal-time hypersurface; here we choose $t'=0$ for the following discussion. With the help of Eq.~(\ref{eq:transform}), we can obtain the side-jump formula either by solving the above condition for $t_*$ or, more intuitively, by shifting the Lorentz-transformed point along its trajectory to the chosen equal-time hypersurface, as shown in Fig.~\ref{fig:lineshift}:
\begin{equation}
\label{eq:jump}
    \boldsymbol\Delta^i = (\bar{\mathbf v}_{\boldsymbol\eta=0}^{\prime i} \Lambda^{0}_{\boldsymbol\eta~j}-\Lambda^{i}_{\boldsymbol\eta~j})[\delta\mathbf{x}+\mathbf{d}]^j_{\boldsymbol\eta=\mathbf 0}+[\delta \mathbf {x}+\mathbf{d}]^i_{\boldsymbol\eta}.
\end{equation}
The $\bar{\mathbf v}^{\prime i}_{\bm\eta=0}$ is defined as $\bar{\mathbf v}^{\prime i}_{\bm\eta=0}=\Lambda^{i}{}_\mu w^\mu/(\Lambda^{0}{}_\mu w^\mu)$ with $w=(\gamma_{\bm\eta=0},\gamma_{\bm\eta=0} \bar{\mathbf v}_{\bm\eta=0})$ and $\gamma_{\bm\eta=0} =1/(1- \bar{\mathbf v}_{\bm\eta=0}^2)^{1/2}$. This transformed velocity $\bar{\mathbf v}^{\prime}_{\bm\eta=0}$  is not necessarily the same as $\bar{\mathbf v}_{\bm\eta}$. The illustration of the formula is shown in Fig.~\ref{fig:lineshift}. 
The contribution from $\mathbf d_{\boldsymbol\eta}$ cancels in the zero-width limit. It also cancels if the expression in square brackets in Eq.~(\ref{eq:deta}) is momentum independent.


\subsection{Isotropic packets and the small-width limit}
In the following, we shall evaluate Eq. (\ref{eq:jump}) in two simple cases, where the squared amplitude $|a(k)|^2$ either depends only on $|\mathbf k|$, or is chosen to be a $\delta$-function centered at $\mathbf p$.

\begin{figure}[!h]
	\includegraphics[width=0.49\textwidth]{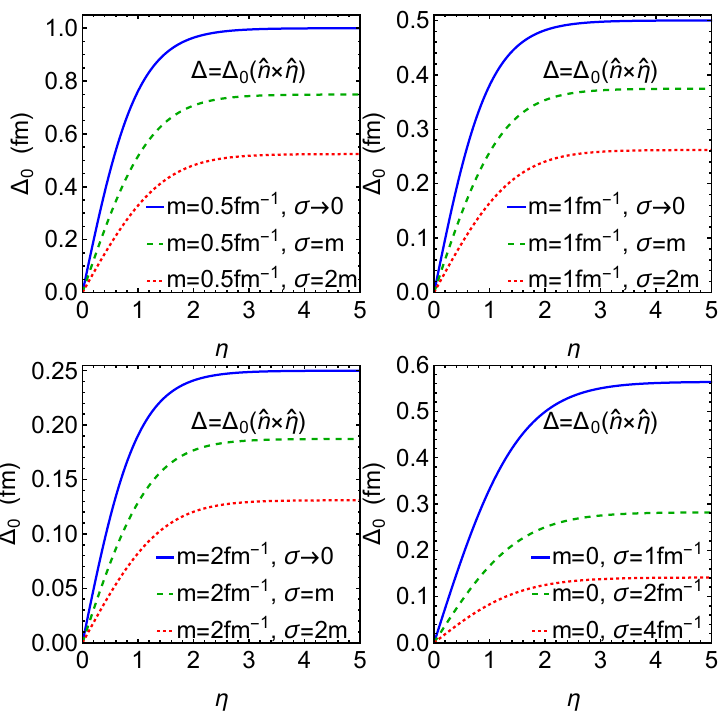}
		\includegraphics[width=0.49\textwidth]{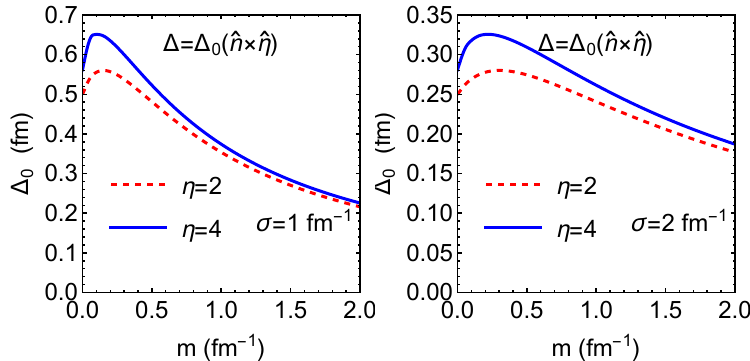}
	\caption{Side jumps under boosts from the rest frame of the wave packet (see Eq.~(\ref{eq:jump1})), obtained using the Gaussian smearing factor with different masses and smearing widths $\sigma$.}
	\label{fig:sidejumps}
\end{figure}
In the case where $|a(k)|^2=|a(|\mathbf k|)|^2$, with a constant phase and a momentum-independent $\xi$, $\bar{\mathbf v}_{\boldsymbol\eta=\mathbf 0}$ vanishes, indicating that the wave-packet center is stationary in the original frame.
This is the case illustrated in Fig. \ref{fig:visualization}. $\delta\mathbf x_{\boldsymbol\eta=\mathbf 0}$ vanishes as well, which makes $\boldsymbol\Delta$ equal to $\delta\mathbf x_{\boldsymbol\eta}$. After dropping the terms whose integrals vanish by symmetry in $\mathbf k$-space in Eq. (\ref{eq:dx}), we obtain
\begin{equation}
    \label{eq:jump1}
    \boldsymbol\Delta=\int \frac{d^3\mathbf k}{(2\pi)^3} |a(k)|^2 \frac{m\sinh \eta-\mathbf k^\prime\cdot\hat{\boldsymbol\eta}}{2(m+\varepsilon_k)\varepsilon_{k^\prime}}(\hat{\mathbf n}\times\hat{\boldsymbol\eta}),
\end{equation}
which is perpendicular to both the polarization direction in the particle rest frame and the boost direction, and agrees with the observation from the visualization shown in Fig.~\ref{fig:visualization}. Note that $\mathbf{k}'\cdot\hat{\eta}=\cosh \eta (\mathbf{k}\cdot \hat{\bm{\eta}})-\sinh \eta\, \varepsilon_k $.
In Fig. \ref{fig:sidejumps}, we plot the side jumps obtained using a Gaussian c-number amplitude $a(|\mathbf k|)$ with different masses and smearing widths $\sigma$.

Based on Fig. \ref{fig:sidejumps}, several features of the side jumps can be identified. First, they saturate for large $\eta$, which is true for an arbitrary choice of $a(|\mathbf k|)$, since, for large $\eta$, $\sinh\eta\approx\cosh\eta$ and the right-hand side of Eq. (\ref{eq:jump1}) approaches
\begin{equation*}
   (\hat{\mathbf n}\times\hat{\boldsymbol\eta})\int \frac{d^3\mathbf k}{(2\pi)^3} |a(k)|^2  \frac{m+\varepsilon_k-\mathbf k\cdot\hat{\boldsymbol\eta}}{2(m+\varepsilon_k)(\varepsilon_k-\mathbf k\cdot\hat{\boldsymbol\eta})},
\end{equation*}
which is independent of $\eta$. Second, under this particular Gaussian choice of $a(|\mathbf k|)$, the side jump  decreases with mass and is proportional to $m^{-1}$ if the ratio $\sigma/m$ is fixed. However, the side jump does not necessarily diverge when $m$ approaches zero. As shown in the middle-right panel of Fig. \ref{fig:sidejumps}, the side jump is finite with a finite smearing width $\sigma$ even though the mass parameter is set equal to zero, which is reasonable, since the total energy of the Gaussian wave-packet in the particle rest frame is finite as long as $\sigma>0$, and increases with the smearing width $\sigma$, indicating that the smearing width plays a similar role to the mass parameter. Consequently, the side jump decreases with the smearing width, as well, which is the third feature of the side jumps.

At fixed smearing width $\sigma$, however, the mass dependence is nonmonotonic, as shown in the lower two panels of Fig. 3. The side jump initially increases from its finite massless value, reaches a maximum, and then decreases at larger masses. Increasing $\sigma$ shifts the maximum to a larger mass, showing that the momentum width sets the mass scale of this behavior. For $m\gg\sigma$, the side jump approaches $\Delta_0\simeq\tanh\eta/(2m)$, recovering the zero-width result.
Meanwhile, it should be noted that a nonzero $\mathbf d_{\boldsymbol\eta}$ can affect the side jump of a finite-width wave packet.

In the case where $|a(k)|^2=(2\pi)^3\delta^3(\mathbf k - \mathbf p)$, $\bar{\mathbf v}_{\boldsymbol\eta=\mathbf 0} = \mathbf p / \varepsilon_{\mathbf p}$ and $\bar{\mathbf v}_{\boldsymbol\eta}=\mathbf p^\prime / \varepsilon_{\mathbf p^\prime}$, indicating that the charge center of the wave-packet moves with the velocities $\mathbf p / \varepsilon_{\mathbf p}$ and $\mathbf p^\prime / \varepsilon_{\mathbf p^\prime}$ in the original and the boosted frame, respectively, which implies $\bar{\mathbf{v}}^{\prime}_{\boldsymbol\eta=0}=\bar{\mathbf{v}}_{\boldsymbol\eta}$. Meanwhile, the contribution from $\mathbf{d}_{\bm\eta}$ cancels completely in this case, so only the $\delta\mathbf{x}$ terms contribute to the side jump. Therefore, we obtain $\boldsymbol\Delta$ and its 4D form $\Delta^\mu$ from Eq. (\ref{eq:jump}) as
\begin{eqnarray}
\label{eq:jump2}
    \boldsymbol\Delta &=& \frac{\sinh\eta}{2 \varepsilon_{p^\prime}}\left[\frac{m}{\varepsilon_p}\hat{\mathbf n}+\left(1-\frac{m}{\varepsilon_p}\right)(\hat{\mathbf n} \cdot \hat{\mathbf p})\hat{\mathbf p}\right]\times\hat{\boldsymbol\eta} \nonumber\\
    &=& \frac{\sinh \eta}{\varepsilon_{p^\prime}\varepsilon_p} \mathbf W\times\hat{\boldsymbol\eta}\\
    \Delta^\mu &=& \frac{\varepsilon^{\mu\alpha\beta\gamma} W^\prime_\alpha u^\prime_\beta \check u_\gamma}{(p^\prime \cdot u^\prime)(p^\prime\cdot\check u)}
\end{eqnarray}
where $\check u=(1,\bm{0})$, $u^\prime \equiv \Lambda \check u=(\cosh\eta,-\sinh\eta\hat{\boldsymbol\eta})$. Also,   $ W^\mu$ is the Pauli-Lubanski vector~\cite{Lubanski:1942A,Lubanski:1942B,Weinberg:1995mt} defined by
\begin{eqnarray}
    W^\mu &\equiv& -\frac {1} {4m} \varepsilon^{\mu\alpha\beta\gamma}\bar u(p)\Sigma_{\alpha\beta}u(p)p_\gamma \nonumber\\
    &=& \frac 1 2 (\hat{\mathbf n}\cdot \mathbf p, m\hat{\mathbf n} + (\varepsilon_p-m)(\hat{\mathbf n} \cdot \hat{\mathbf p})\hat{\mathbf p})
\end{eqnarray}
which is covariant and equal to $(0,m\hat{\mathbf n}/2)$ in the particle rest frame, with $\Sigma_{\alpha\beta}$ being the spin tensor operator, defined as $(i/4)[\gamma_\alpha,\gamma_\beta]$ for the Dirac spinor. In the massless chiral limit, $\mathbf W = \lambda \mathbf p$~\cite{Weinberg:1995mt} where $\lambda$ is the helicity of the particle. Thus, Eq. (\ref{eq:jump2}) gives the side jump coinciding with the one written in Ref~\cite{Chen:2015gta}. Based on the covariance of the total angular momentum, we can also derive Eq. (\ref{eq:jump2}), which is briefly discussed in  App.~\ref{sec_cons}.
\section{Summary}
\label{sec_sum}

In this work, we have developed a trace method to derive the anomalous Lorentz transformation of the center of the probability density of a free Dirac wave packet. The method gives a compact analytical expression for the spin-dependent displacement under a finite boost, while retaining the momentum spread of the packet and including a momentum-dependent spin direction, without a semiclassical expansion. The side jump is defined on the equal-time hypersurface relative to the Lorentz-transformed original center trajectory.

For the spherically symmetric packets considered here, the side jump is perpendicular to both the initial spin direction and the boost direction. For the Gaussian packets considered here, it is inversely proportional to the mass at fixed $\sigma/m$ and remains finite in the massless limit at fixed nonzero $\sigma$. At fixed momentum width, the side jump instead increases at small masses before decreasing at larger masses. This nonmonotonic behavior shows that the finite momentum spread modifies the mass dependence beyond the $1/m$ scaling of the zero-width result. In the small width limit, a compact result expressed in terms of the Pauli–Lubanski vector has been obtained and agrees with the displacement required by the Lorentz covariance of the total angular momentum. For states of definite helicity, it also reproduces the known massless result.

The trace method and the resulting side-jump formula provide a basis for further studies. The side jumps given either by Eq. (\ref{eq:jump2}) or equivalently by Eq. (\ref{eq:CovaraintJump}) can be implemented in the transport models where spins are treated as the dynamical degrees of freedom, like positions and momenta. Such transport models can be employed to investigate various intriguing phenomena, including the spin polarization in the heavy-ion collisions, which we leave for future work. Meanwhile, it would also be interesting to investigate the connection between the side-jump formulation proposed in this work and the other theoretical approaches concerning the related topic~\cite{Florkowski:2017ruc,Becattini:2018duy,Weickgenannt:2019dks,Gao:2019znl,Yang:2020hri,Bhadury:2020puc}.

Jin-Rui Wu and Shuai Liu are supported by NSFC No. 12205090 and Fundamental Research Funds for the central Universities. Feng Li is supported by NSFC No. 12105129.

\appendix
\section{Details of the derivations}
\label{sec_details}
In this section, we discuss the details of the derivation to obtain Eq.~(\ref{eq:dx}).

\subsection{Vanishing of the imaginary part of $\d \mathbf{x}$}
It should be noted that in Eq.~(\ref{eq:x2}), there are imaginary terms in the integrand contributing to $\d \vx$. In the following, we will show that these terms can be organized into a total derivative whose integral vanishes under the boundary condition.

There are generally two sources of these contributions: derivatives of the energy factors and of $a_0(k)$ from Eq.~(\ref{eq:x2}), and  a contribution from $X$ in Eq.~(\ref{eq:deltax}). 
First, we will illustrate the contributions from the derivatives of the energy factors and of $a_0(k)$. With the help of the identity Eq.~(\ref{eq:ussu}),
these contributions can be expressed as
\begin{align}
	\text{term}_\ve&=\frac{a^\ast(k)}{\sqrt{2\varepsilon_k}}\partial_{\mathbf k^\prime}\left(\frac{\sqrt{\varepsilon_k}}{\sqrt{2}\varepsilon_{k^\prime}} \right)a(k)(2\ve_k')\no\\
	&=(\frac{1}{2} \frac{\pd_{\vk'}\ve_{k}}{\ve_{k}}-\frac{\pd_{\vk'}\ve_{k'}}{\ve_{k'}})|a(k)|^2 \\
	\text{term}_a&=\frac{a_0(k)}{\sqrt{2\varepsilon_k}}\frac{\varepsilon_k}{\varepsilon_{k^\prime}}\frac{\partial_{\mathbf k^\prime}a_0(k)}{\sqrt{2\varepsilon_k}}2\ve_k'=\frac{1}{2}\pd_{\vk'}|a(k)|^2
\end{align}
The $X$ contribution comes from the “1” part in the second line of Eq.~(\ref{eq:tracetech}) and can be expressed as
\begin{align}
\text{term}_{X}&=\frac{1}{2\ve_{k'}}\frac{m}{2}\text{Tr}\left\{
	\slashed{h}_{\bm\eta_{k}}\slashed{h}_{\bm\eta}\slashed{h}_{-\bm\eta} \partial_{\mathbf k^\prime}\slashed{h}_{-\bm\eta_{k}}\right\}|a(k)|^2\no\\
	&=\frac{1}{2\ve_{k'}}\frac{m}{2}\frac{1}{2}\pd_{\vk'} [\text{Tr}\left\{
	\slashed{h}_{\bm\eta_{k}}\slashed{h}_{\bm\eta}\slashed{h}_{-\bm\eta} \slashed{h}_{-\bm\eta_{k}}\right\}]|a(k)|^2\no\\	
	&=\frac{1}{2\ve_{k'}}\frac{m}{2}\frac{1}{2}\pd_{\vk'} [\frac{4\ve_{k'}}{m}]|a(k)|^2=\frac{\partial_{\mathbf k^\prime}\varepsilon_{k'}}{2\ve_{k'}}|a(k)|^2
\end{align}
The total derivative can be identified after we evaluate the trace and note that $h^\mu_{\bm\eta_{k}}\cdot\pd(h_{-\bm\eta_{k},\mu})=(1/2)\pd(h_{\bm\eta_{k}}\cdot h_{-\bm\eta_{k}})$. To obtain the third line above, we use the identity in Eq.~(\ref{eq:ussu}).

Summing all the  imaginary terms, we obtain:
\begin{align}
	&i\int \frac{d^3\mathbf k}{(2\pi)^3}\left[\text{term}_{\ve}+\text{term}_{a}+\text{term}_{X} \right]\no\\
	&=\frac{i}{2}\int \frac{d^3\mathbf k'}{(2\pi)^3}[(\frac{\partial_{\mathbf k^\prime}\ve_{k}}{\ve_{k'}}-\frac{\ve_{k}\partial_{\mathbf k^\prime}\ve_{k'}}{\ve_{k'}^2})|a(k)|^2+\frac{\ve_{k}}{\ve_{k^\prime}}\partial_{\mathbf k^\prime}|a(k)|^2]\no\\
	&=\frac{i}{2}\int \frac{d^3\mathbf k'}{(2\pi)^3}\partial_{\mathbf k^\prime}[\frac{\ve_{k}}{\ve_{k'}}|a(k)|^2]=0
\end{align}

\subsection{Details of the real part calculation}
In this section, we discuss the details of the derivation of Eq.~(\ref{eq:dx}) from the second term of $X(k)$, given in Eq.~(\ref{eq:eta epsilon}). Before we start, we prepare some essential components that are helpful in simplifying the following derivation:
\begin{align}
	\label{eq:keyterms}
	&\partial_{j} h_{-\bm\eta_k,\beta}=\left[ \left( \bar{\delta}_{\b j}+\dfrac{\bar{k}_\b \bar{k}_{j}}{2\varepsilon_k(\varepsilon_k+m)}\right) -\frac{\check{u}_{\b}\bar{k}_{j}}{2\varepsilon_{k}} \right] D^{-1}_{h_k},\\
	&h_{\bm\eta_{k},\r}
	=[-\bar{k}_{\r}+(m+\varepsilon_{ k})\check{u}_{\r}]D^{-1}_{h_k},\\
	&h_{\bm{\eta},\s}
	=[\bar{u}_{\s}+(1+u^{0})\check{u}_{\s}]D^{-1}_{u},\\
	&h_{-\bm\eta,\a}
	=[ -\bar{u}_{\a}+(1+u^{0})\check{u}_{\a}]D^{-1}_{u}.
\end{align}
Here, $D_{h_k}=\sqrt{2m(\ve_k+m)}$ and $D_{u}=\sqrt{2(1+u^0)}$. The $\partial_{j}$ denotes  differentiation with respect to $\bar{k}^j$. We factor out the common prefactor from $X_2$ and define $Y^{(n)}_j$ corresponding to the six terms in  Eq.~(\ref{eq:eta epsilon}). Although the following derivation uses the above representation, we also note that $\partial_{j} h_{-\bm\eta_k,\beta}=\big[\bar{\delta}_{\b j}-{\check{u}_{\b}\bar{k}_{j}}/{\varepsilon_{k}} +c  h_{-\bm\eta_k,\beta}\bar{k}_j\big]D^{-1}_{h_k}$, where the $c  h_{-\bm\eta_k,\beta}\bar{k}_j D^{-1}_{h_k}$ term will not contribute after summing up all contributions, as expected. Thus, we can take $\partial_{j} h_{-\bm\eta_k,\beta}=(\bar{\delta}_{\b j}-{\check{u}_{\b}\bar{k}_{j}}/{\varepsilon_{k}} )D^{-1}_{h_k}$ in the calculation to obtain the same results. We now write $X_2$ as
\begin{align}\label{XtoY}
	X_2^{i'}=\frac{i(2m)}{D^{2}_{u}D^{2}_{h_k}} \sum_{n}(-\bar{\d}^{ii'}) Y^{(n)}_j M^{j}{}_{i},
\end{align}
where $M^{j}{}_{i}$ is the Jacobian matrix $\partial\bar{k}^j/\partial\bar{k}'^i$, and the index $n$ in $Y^{(n)}_j$ follows the order in Eq.~(\ref{eq:eta epsilon}). The first term, $Y^{(1)}_j$, can be expressed as
\begin{align}\label{eq:Y1}
	Y^{(1)}_j=&
	\eta^{\alpha\beta}\epsilon^{\mu\nu\rho\sigma}\check{u}_{\m}
	 n_\nu h_{\bm\eta_k,\rho}h_{\bm\eta,\sigma}
	h_{-\bm\eta,\alpha}\partial_{j} h_{-\bm\eta_k,\beta}D_{\text{all}}\no\\
	=&\left[ -\frac{(1+u^0)\bar k_j}{2\ve_k}-\bar{u}_j-\frac{(\bar{k}\cdot \bar{u})\bar k_j}{2\ve_k(\ve_k+m)}\right] \no\\
	&\times\epsilon^{\m\nu\rho\sigma}\check{u}_{\m}n_{\nu}( - \bar{k}_{\r})\bar{u}_{\s},
\end{align}
where the expression $D_{\text{all}}=D^2_uD^2_{h_k}$ is a shorthand for the denominators.
The expression can be obtained by carefully applying the expressions in Eq.~(\ref{eq:keyterms}). Similarly, the result for the second term $Y^{(2)}_j$ is
\begin{align}\label{eq:Y2}
	Y^{(2)}_j=&
	\eta^{\m\b}\epsilon^{\a\nu\rho\sigma}\check{u}_{\m}
	n_\nu h_{\bm\eta_k,\rho}h_{\bm\eta,\sigma}
	h_{-\bm\eta,\alpha}\partial_{j} h_{-\bm\eta_k,\beta}D_{\text{all}}\no\\
	=&-\frac{(1+u^0)\bar{k}_{j}}{\varepsilon_{k}}\epsilon^{\a\nu\rho\sigma}\check{u}_{\a}n_{\nu}( - \bar{k}_{\r})  \bar{u}_{\s}
\end{align}
For the third term $Y^{(3)}_j$, there are four sub-terms, which we label $Y^{(3,n)}_j$ and which are as follows:
\begin{align}\label{eq:Y3}
	Y^{(3)}_j&=
	-\eta^{\a\m}\epsilon^{\b\nu\rho\sigma}
	\check{u}_\m n_\nu h_{\bm\eta_k,\rho}h_{\bm\eta,\sigma}
	h_{-\bm\eta,\alpha}\partial_{j} h_{-\bm\eta_k,\beta}D_{\text{all}}\no\\
	&=Y^{(3,1)}_j+Y^{(3,2)}_j+Y^{(3,3)}_j+Y^{(3,4)}_j;\\
		&\hspace{-1.8em}
		Y^{(3,1)}_j=\frac{(1+u^0)\bar{k}_{j}}{2\varepsilon_{k}}\epsilon^{\b\nu\rho\sigma}\check{u}_{\b}n_{\nu}( - \bar{k}_{\r})\bar{u}_{\s},\no\\
		&\hspace{-1.8em}
		Y^{(3,2)}_j=-(1+u^0)(m+\ve_{k})\epsilon_{j}^{\;\;\nu\r\sigma}n_{\nu}\check{u}_{\r}\bar{u}_{\s},\no\\
		&\hspace{-1.8em}
		Y^{(3,3)}_j=-\frac{(1+u^0)\bar{k}_{j}}{2\varepsilon_{k}}\epsilon^{\b\nu\r\sigma}\bar{k}_{\b}n_{\nu}\check{u}_{\r}\bar{u}_{\s},\no\\
		&\hspace{-1.8em}
		Y^{(3,4)}_j=-(1+u^0)^2
		\epsilon_{j}^{\;\;\nu\r\s}n_{\nu}(-\bar{k}_{\r})\check{u}_{\s}.\no
\end{align}
It should be noted that $Y^{(3,1)}_j$ and $Y^{(3,3)}_j$ cancel each other exactly. Thus, only the two terms $Y^{(3,2)}_j$
and $Y^{(3,4)}_j$ will contribute to $Y^{(3)}_j$. For 	$Y^{(4)}_j$, it is not hard to see that it vanishes. For 	$Y^{(5)}_j$, there are two sub-terms:
\begin{align}\label{eq:Y5}
	&Y^{(5)}_j=
	-\eta^{\s\r}\epsilon^{\n\a\b\m}\check{u}_{\m}
	n_\nu h_{\bm\eta_k,\rho}h_{\bm\eta,\sigma}
	h_{-\bm\eta,\alpha}\partial_{j} h_{-\bm\eta_k,\beta}D_{\text{all}}\no\\
	&\qquad=Y^{(5,1)}_{j}+Y^{(5,2)}_{j};\\
	&Y^{(5,1)}_{j}=[\bar{k}\cdot \bar{u} -(1+u^0)(m+\ve_k)]\epsilon_{j}^{\;\;\n\a\m}\check{u}_\m n_{\n}(-\bar{u}_{\a}),\no\\
	&Y^{(5,2)}_{j}=\left[\frac{(\bar{k}\cdot \bar{u})\bar{k}_j}{2\ve_k(\ve_k+m)} -\frac{(1+u^0)\bar{k}_j}{2\ve_{k}}\right]\epsilon^{\n\a\b\m}\check{u}_\m n_{\n}(-\bar{u}_{\a})\bar{k}_\b.
    \no 
\end{align}
Finally, $Y^{(6)}_j$ can be expressed as follows:
\begin{align}\label{eq:Y6}
	Y^{(6)}_j=&
	\eta^{\n\s}\epsilon^{\r\a\b\m}\check{u}_{\m}
	n_\nu h_{\bm\eta_k,\rho}h_{\bm\eta,\sigma}
	h_{-\bm\eta,\alpha}\partial_{j} h_{-\bm\eta_k,\beta}D_{\text{all}}\no\\
	=&(n\cdot \bar{u})\epsilon_{j}^{\;\;\r\a\m}\check{u}_\m(-\bar{u}_{\a})(-\bar{k}_{\r}),
\end{align}
and this structure will be shown to be absorbed into other contributions after the Lorentz transformation.

All terms in the $Y^{(n)}$'s can be organized by their tensor structures into four terms:
\begin{align}
	&Y^{(a)}_j=Y^{(1)}_j+Y^{(2)}_j+Y^{(5,2)}_j\no\\
	&\hspace{1.8em}=\left[ 2(1+u^0)\frac{\bar k_j}{\ve_{k}}+\bar{u}_j\right] \ve^{\a\n\r\s}\check{u}_{\a}n_{\n}\bar{k}_{\r}\bar{u}_{\s},\\
	&Y^{(b)}_j=Y^{(3,2)}_j+Y^{(5,1)}_j\no\\
	&\hspace{1.8em}=[2(1+u^0)(m+\ve_k)-\bar{k}\cdot\bar{u}]\ve_{j}{}^{\n\a\m}n_{\n}\bar{u}_{\a}\check{u}_{\m},\\
	&Y^{(c)}_j=Y^{(3,4)}_j=(1+u^0)^2\ve_{j}{}^{\n\r\s}n_{\n}\bar{k}_{\r}\check{u}_{\s},\\
	&Y^{(d)}_j=Y^{(6)}_j=-(n\cdot \bar{u})\ve_{j}{}^{\a\r\m}\bar{u}_{\a}\bar{k}_{\r}\check{u}_{\m}.
\end{align}
It remains to perform the boost, for which the Jacobian is
\begin{align}
	M^{j}{}_{i}&=[\Lambda_{-\bm\eta}]^{j}_{\;i} -[\Lambda_{-\bm\eta}]^j_{\;0}\frac{\bar{k}'_{ i}}{\varepsilon_{k'}}\no\\
	&=\bar{\d}^j{}_i-(\cosh\eta-1)\hat{\bar{\eta}}^j\hat{\bar{\eta}}_i-\sinh\eta \hat{\bar{\eta}}^j\frac{\bar{k}'_i}{\ve_{k'}}\\
	&=\bar{\d}^j{}_i+[M_{u}]^{j}{}_i,\;\; [M_{u}]^{j}{}_i=-\frac{\bar u^j\bar u_i}{1+u^0}+\bar u^j\frac{\bar{k}_i'}{\ve_{k'}}.
\end{align}
Note that $\bar{k}'_i=\bar{k}_i-\bar{u}_i{\bar{k}\cdot \bar{u}}/{(1+u^0)}+\bar{u}_i\ve_{k}$, $\bar{u}^i=-\sinh\eta \hat{\bar{\eta}}^i$, and $\ve_{k'}=u^0\ve_{k}-\bar{u}\cdot \bar k$. The scalar products of the spatially projected four-vectors have the opposite sign to the corresponding three-vector dot products.
After transformation with $M^{j}_i$, we obtain
\begin{align}\label{JY1}
	&[M\cdot Y]^{(1)}_i=[M\cdot Y^{(a)}+M_u\cdot Y^{(c)}]_i\no\\
	&=\left[\left(2(1+u^0)\frac{\bar k_i}{\ve_{k}}+\bar{u}_i\right) -\left( 2\frac{\bar k\cdot \bar u}{\ve_{k}}+1-u^0\right) \bar{u}_i\right.\no\\
	&\quad\left.+\left( 2(1+u^0)\frac{\bar k\cdot \bar u}{\ve_{k}}+1-(u^0)^2\right) \frac{\bar{k}'_i}{\ve_{k'}}+(1+u^0)\bar{u}_i\right.\no\\
	&\quad\left.-(1+u^0)^2 \frac{\bar{k}'_i}{\ve_{k'}}\right]\ve^{\a\n\r\s}\check{u}_{\a}n_{\n}\bar{k}_{\r}\bar{u}_{\s}\no\\
	&=-\bar{u}_i \ve^{\a\n\r\s}\check{u}_{\a}n_{\n}\bar{k}_{\r}\bar{u}_{\s}.
\end{align}
The other three terms are simpler:
\begin{align}
&[M\cdot Y ]^{(2)}_i=\left[ 2(1+u^0)(m+\ve_k)\right.\no\\
&\hspace{6em}\left.-\bar{k}\cdot\bar{u}\right]\ve_{i}{}^{\n\a\m}n_{\n}\bar{u}_{\a}\check{u}_{\m},\\
&[M\cdot Y ]^{(3)}_i=(1+u^0)^2\ve_{i}{}^{\n\r\s}n_{\n}\bar{k}_{\r}\check{u}_{\s},\\
&[M\cdot Y]^{(4)}_i=-(n\cdot \bar{u})\ve_{i}{}^{\a\r\m}\bar{u}_{\a}\bar{k}_{\r}\check{u}_{\m}\label{JY4}.
\end{align}
Combining  Eqs.~(\ref{JY1})--(\ref{JY4}), we can express the result in terms of three-dimensional vectors:
\begin{align}
	[\bm{M}\cdot\bm{Y}]&=[2(1+u^0)(m+\epsilon_k)+\mathbf{k} \cdot \mathbf{u}]\mathbf{n} \times \mathbf{u}\nonumber\\
	&\quad+(1+u^0)^2\mathbf{n}\times\mathbf{k}+(\mathbf{n}\cdot \mathbf{u})(\mathbf{u}\times\mathbf{k})\nonumber\\
	&\quad-\mathbf{u}\mathbf{u}\cdot(\mathbf{n}\times\mathbf{k}).
\end{align}
Note that the vectors satisfy, for example, $\bar{k}_i=-\mathbf{k}_i$ and $\bar{k}\cdot\bar{u}=-\mathbf{k} \cdot \mathbf{u}$, where bold symbols follow three-dimensional Euclidean conventions. As the expression shows, the final results contain five different vectors. However, there are only three independent directions in three-dimensional space. We employ the identity
\begin{align}
	&(\mathbf{u}\cdot\mathbf{k})(\mathbf{n}\times\mathbf{u})-(\mathbf{u}\cdot\mathbf{n})(\mathbf{k}\times\mathbf{u})=(\mathbf{u}\times(\mathbf{n}\times\mathbf{k}))\times \mathbf{u}\no\\
	&\qquad=(\mathbf{u}\cdot\mathbf{u})(\mathbf{n}\times\mathbf{k})-(\mathbf{n}\times\mathbf{k})\cdot\mathbf{u}\mathbf{u},
\end{align}
Using this identity, we obtain
\begin{align}
	[\bm{M}\cdot\bm{Y}]=&2(1+u^0)(m+\epsilon_k)(\mathbf{n}\times\mathbf{u})+2u^0(1+u^0)\mathbf{n}\times\mathbf{k}\nonumber\\
	&-2(\mathbf{n}\times\mathbf{k})\cdot\mathbf{u}\mathbf{u}
\end{align}
Employing $\mathbf{u}=-\sinh\eta \hat{\boldsymbol\eta}$ and evaluating Eq. (\ref{XtoY}), we obtain 
\begin{align}
	\bm X_2&=i\left[-\frac{\cosh\eta-1}{m+\epsilon_k}(\mathbf{n}\times\mathbf{k})\cdot\hat{\bm\eta}\hat{\bm\eta}+ \frac{\cosh\eta}{m+\epsilon_k}\mathbf{n}\times\mathbf{k}\right.
	\no\\
	&\qquad\qquad\quad\left.-\sinh\eta(\mathbf{n}\times\hat{\bm\eta})\right],
\end{align}
from which one can obtain the explicit expression for $\delta\mathbf{ x}_{\bm\eta}$, i.e., Eq. (\ref{eq:dx}).

\section{Derivation from covariance of $J^{\mu\nu}$}
\label{sec_cons}
The total angular momentum $J^{\mu\nu}$ is the conserved charge corresponding to the $SO(1,3)$ symmetry, and is therefore equal to
\begin{equation}
	\label{eq:Jmunu}
	\int d^3\mathbf x\mathcal{J}^{0\mu\nu},
\end{equation}
where $\mathcal{J}$ is the corresponding Noether current, obtained using the conventional Lagrangian for the Dirac spinor~\cite{Peskin:1995ev} and given by
\begin{equation}
	\label{eq:Jcurrent}
	\mathcal{J}^{\alpha\mu\nu}=\bar{\psi}i\gamma^\alpha(x^\mu\partial^\nu-x^\nu\partial^\mu-i\Sigma^{\mu\nu})\psi.
\end{equation}
The integration in Eq. (\ref{eq:Jmunu}) is again evaluated under the assumption that $|a(k)|^2$ is a delta function centered at $\mathbf p$, and $J^{\mu\nu}$ is thus expressed as a sum of the orbital and the spin components, i.e.,
\begin{equation}
	J^{\mu\nu}=L^{\mu\nu}+S^{\mu\nu}
\end{equation}
where 
\begin{eqnarray}
	L^{\mu\nu}=\bar x^\mu p^\nu-\bar x^\nu p^\mu,\quad S^{\mu\nu}=\frac{\varepsilon^{\mu\nu\alpha\beta} W_\alpha \check{u}_\beta}{p\cdot \check{u}}
	\label{eq:Smunu}
\end{eqnarray}
with $\check{u}$ being $(1,\mathbf 0)$ in all the frames and therefore not a covariant vector. It needs to be clarified that both $L^{\mu\nu}$ and $S^{\mu\nu}$ represent only the real parts of the expectation values of $i(x^\mu\partial^\nu-x^\nu\partial^\mu)$ and $\Sigma^{\mu\nu}$, respectively, while the imaginary parts of the latter two expectation values are equal in magnitude but opposite in sign, and therefore cancel each other in Eq. (\ref{eq:Jcurrent}). Also, the decomposition of $J^{\mu\nu}$ is not unique; it can also be changed by  a pseudo-gauge transformation~\cite{Becattini:2018duy}.

The Lorentz covariance of $S^{\mu\nu}$ is explicitly broken by $\check u$ in Eq.  (\ref{eq:Smunu}), while the total angular momentum $J^{\mu\nu}$ should be covariant. We therefore construct an anomalous Lorentz transformation for $\bar x$, under which $\bar x$ transforms as $\bar x \to \bar x^\prime + \Delta$, so that the covariance of the total angular momentum $J^{\mu\nu}$ is preserved. In the following discussion, any primed 4-vector, such as $a^\prime$, stands for a new 4-vector equal to the original one multiplied by the Lorentz transformation matrix, such as $\Lambda a$, regardless of whether the 4-vector is covariant or not, so $\bar x^\prime$ stands for $\Lambda \bar x$. Under a boost, the total angular momentum $J^{\mu\nu}$ transforms as
\begin{equation}
	\label{eq:transJ1}
	J^{\mu\nu} \to (\bar x^{\prime\mu} + \Delta^\mu) p^{\prime\nu}-(\bar x^{\prime\nu}+\Delta^\nu) p^{\prime\mu} + \frac{\varepsilon^{\mu\nu\alpha\beta} W^\prime_\alpha \check{u}_\beta}{p^\prime \cdot \check{u}}.
\end{equation}
Meanwhile, the covariance of $J^{\mu\nu}$ can be made explicit if we pretend that both $\bar x$ and $\check u$ are covariant and  transform as $\bar x \to \bar x^\prime$ and $\check u \to u^\prime \equiv \Lambda \check u=(\cosh\eta,-\sinh\eta\hat{\boldsymbol\eta})$, respectively. Under such a hypothesis, the total angular momentum $J^{\mu\nu}$ should transform as
\begin{equation}
	\label{eq:transJ2}
	J^{\mu\nu} \to \bar x^{\prime\mu} p^{\prime\nu}-\bar x^{\prime\nu} p^{\prime\mu} + \frac{\varepsilon^{\mu\nu\alpha\beta} W^\prime_\alpha u^\prime_\beta}{p^\prime \cdot u^\prime}.
\end{equation}
We learn, by comparing Eq. (\ref{eq:transJ1}) with Eq. (\ref{eq:transJ2}), that the total angular momentum $J^{\mu\nu}$ is covariant as long as the anomalous displacement $\Delta$ satisfies
\begin{equation}
	\label{eq:conditionD}
	\Delta^\mu p^{\prime\nu}-\Delta^\nu p^{\prime\mu}=\frac{\varepsilon^{\mu\nu\alpha\beta} W^\prime_\alpha u^\prime_\beta}{p^\prime \cdot u^\prime}-\frac{\varepsilon^{\mu\nu\alpha\beta} W^\prime_\alpha \check{u}_\beta}{p^\prime \cdot \check{u}}.
\end{equation}
We therefore obtain $\Delta$, which is again assumed to be purely spatial, by contracting both sides of Eq.~(\ref{eq:conditionD}) with $\check u_\nu$, giving
\begin{equation}
	\label{eq:CovaraintJump}
	\Delta^\mu = \frac{\varepsilon^{\mu\alpha\beta\gamma} W^\prime_\alpha u^\prime_\beta \check u_\gamma}{(p^\prime \cdot u^\prime)(p^\prime\cdot\check u)},
\end{equation}
which, after the substitution $W^\prime \to \lambda p^\prime$ corresponding to the massless chiral limit, is exactly the same side jump as the one provided in Ref.~\cite{Chen:2015gta}. The obtained side jump also contributes to the particle number current in the same way as demonstrated in Ref.~\cite{Chen:2015gta}, with the $\Delta$ in the literature being replaced with Eq. (\ref{eq:CovaraintJump}), and leads to anomalous currents. Given the identity $\mathbf W^\prime \times \hat{\boldsymbol\eta}=\mathbf W\times \hat{\boldsymbol\eta}$, it is easy to show that the side jumps obtained in both approaches, i.e.,  Eq. (\ref{eq:jump2}) and Eq. (\ref{eq:CovaraintJump}), are identical. Therefore, we have proved that the covariance of the total angular momentum is preserved when the anomalous Lorentz transformation with the side jumps given either by Eq. (\ref{eq:jump2}) or by Eq. (\ref{eq:CovaraintJump}) is employed, and the angular momentum conservation law, which would in general be broken under a normal Lorentz transformation, is thus restored. It should be mentioned that the same results were obtained using classical field theory in Ref.~\cite{Lorce:2018zpf}, while similar physics are discussed in Ref.~\cite{Hattori:2019ahi}. However, this work derives the formula from full quantum theory, with a natural extension to finite-width wave packets.
Furthermore, since the spin representation is not specified in the definition of the Pauli-Lubanski vector, both Eq. (\ref{eq:jump2}) and Eq. (\ref{eq:CovaraintJump}) might hold not only for the spin-half fermions but also for any spinning particles, which is worth checking in the future.
\bibliography{ref}

\begin{thebibliography}{34}%
\makeatletter
\providecommand \@ifxundefined [1]{%
 \@ifx{#1\undefined}
}%
\providecommand \@ifnum [1]{%
 \ifnum #1\expandafter \@firstoftwo
 \else \expandafter \@secondoftwo
 \fi
}%
\providecommand \@ifx [1]{%
 \ifx #1\expandafter \@firstoftwo
 \else \expandafter \@secondoftwo
 \fi
}%
\providecommand \natexlab [1]{#1}%
\providecommand \enquote  [1]{``#1''}%
\providecommand \bibnamefont  [1]{#1}%
\providecommand \bibfnamefont [1]{#1}%
\providecommand \citenamefont [1]{#1}%
\providecommand \href@noop [0]{\@secondoftwo}%
\providecommand \href [0]{\begingroup \@sanitize@url \@href}%
\providecommand \@href[1]{\@@startlink{#1}\@@href}%
\providecommand \@@href[1]{\endgroup#1\@@endlink}%
\providecommand \@sanitize@url [0]{\catcode `\\12\catcode `\$12\catcode
  `\&12\catcode `\#12\catcode `\^12\catcode `\_12\catcode `\%12\relax}%
\providecommand \@@startlink[1]{}%
\providecommand \@@endlink[0]{}%
\providecommand \url  [0]{\begingroup\@sanitize@url \@url }%
\providecommand \@url [1]{\endgroup\@href {#1}{\urlprefix }}%
\providecommand \urlprefix  [0]{URL }%
\providecommand \Eprint [0]{\href }%
\providecommand \doibase [0]{https://doi.org/}%
\providecommand \selectlanguage [0]{\@gobble}%
\providecommand \bibinfo  [0]{\@secondoftwo}%
\providecommand \bibfield  [0]{\@secondoftwo}%
\providecommand \translation [1]{[#1]}%
\providecommand \BibitemOpen [0]{}%
\providecommand \bibitemStop [0]{}%
\providecommand \bibitemNoStop [0]{.\EOS\space}%
\providecommand \EOS [0]{\spacefactor3000\relax}%
\providecommand \BibitemShut  [1]{\csname bibitem#1\endcsname}%
\let\auto@bib@innerbib\@empty
\bibitem [{\citenamefont {Berger}(1970)}]{PhysRevB.2.4559}%
  \BibitemOpen
  \bibfield  {author} {\bibinfo {author} {\bibfnamefont {L.}~\bibnamefont
  {Berger}},\ }\bibfield  {title} {\bibinfo {title} {Side-jump mechanism for
  the hall effect of ferromagnets},\ }\href@noop {} {\bibfield  {journal}
  {\bibinfo  {journal} {Phys. Rev. B}\ }\textbf {\bibinfo {volume} {2}},\
  \bibinfo {pages} {4559} (\bibinfo {year} {1970})}\BibitemShut {NoStop}%
\bibitem [{\citenamefont {Nagaosa}\ \emph {et~al.}(2010)\citenamefont
  {Nagaosa}, \citenamefont {Sinova}, \citenamefont {Onoda}, \citenamefont
  {MacDonald},\ and\ \citenamefont {Ong}}]{RevModPhys.82.1539}%
  \BibitemOpen
  \bibfield  {author} {\bibinfo {author} {\bibfnamefont {N.}~\bibnamefont
  {Nagaosa}}, \bibinfo {author} {\bibfnamefont {J.}~\bibnamefont {Sinova}},
  \bibinfo {author} {\bibfnamefont {S.}~\bibnamefont {Onoda}}, \bibinfo
  {author} {\bibfnamefont {A.~H.}\ \bibnamefont {MacDonald}},\ and\ \bibinfo
  {author} {\bibfnamefont {N.~P.}\ \bibnamefont {Ong}},\ }\bibfield  {title}
  {\bibinfo {title} {Anomalous hall effect},\ }\href
  {https://doi.org/10.1103/RevModPhys.82.1539} {\bibfield  {journal} {\bibinfo
  {journal} {Rev. Mod. Phys.}\ }\textbf {\bibinfo {volume} {82}},\ \bibinfo
  {pages} {1539} (\bibinfo {year} {2010})}\BibitemShut {NoStop}%
\bibitem [{\citenamefont {Sinova}\ \emph {et~al.}(2015)\citenamefont {Sinova},
  \citenamefont {Valenzuela}, \citenamefont {Wunderlich}, \citenamefont
  {Back},\ and\ \citenamefont {Jungwirth}}]{RevModPhys.87.1213}%
  \BibitemOpen
  \bibfield  {author} {\bibinfo {author} {\bibfnamefont {J.}~\bibnamefont
  {Sinova}}, \bibinfo {author} {\bibfnamefont {S.~O.}\ \bibnamefont
  {Valenzuela}}, \bibinfo {author} {\bibfnamefont {J.}~\bibnamefont
  {Wunderlich}}, \bibinfo {author} {\bibfnamefont {C.~H.}\ \bibnamefont
  {Back}},\ and\ \bibinfo {author} {\bibfnamefont {T.}~\bibnamefont
  {Jungwirth}},\ }\bibfield  {title} {\bibinfo {title} {Spin hall effects},\
  }\href {https://doi.org/10.1103/RevModPhys.87.1213} {\bibfield  {journal}
  {\bibinfo  {journal} {Rev. Mod. Phys.}\ }\textbf {\bibinfo {volume} {87}},\
  \bibinfo {pages} {1213} (\bibinfo {year} {2015})}\BibitemShut {NoStop}%
\bibitem [{\citenamefont {Chen}\ \emph {et~al.}(2014)\citenamefont {Chen},
  \citenamefont {Son}, \citenamefont {Stephanov}, \citenamefont {Yee},\ and\
  \citenamefont {Yin}}]{Chen:2014cla}%
  \BibitemOpen
  \bibfield  {author} {\bibinfo {author} {\bibfnamefont {J.-Y.}\ \bibnamefont
  {Chen}}, \bibinfo {author} {\bibfnamefont {D.~T.}\ \bibnamefont {Son}},
  \bibinfo {author} {\bibfnamefont {M.~A.}\ \bibnamefont {Stephanov}}, \bibinfo
  {author} {\bibfnamefont {H.-U.}\ \bibnamefont {Yee}},\ and\ \bibinfo {author}
  {\bibfnamefont {Y.}~\bibnamefont {Yin}},\ }\bibfield  {title} {\bibinfo
  {title} {{Lorentz Invariance in Chiral Kinetic Theory}},\ }\href
  {https://doi.org/10.1103/PhysRevLett.113.182302} {\bibfield  {journal}
  {\bibinfo  {journal} {Phys. Rev. Lett.}\ }\textbf {\bibinfo {volume} {113}},\
  \bibinfo {pages} {182302} (\bibinfo {year} {2014})}\BibitemShut {NoStop}%
\bibitem [{\citenamefont {Duval}\ \emph {et~al.}(2015)\citenamefont {Duval},
  \citenamefont {Elbistan}, \citenamefont {Horváthy},\ and\ \citenamefont
  {Zhang}}]{Duval:2014cfa}%
  \BibitemOpen
  \bibfield  {author} {\bibinfo {author} {\bibfnamefont {C.}~\bibnamefont
  {Duval}}, \bibinfo {author} {\bibfnamefont {M.}~\bibnamefont {Elbistan}},
  \bibinfo {author} {\bibfnamefont {P.~A.}\ \bibnamefont {Horváthy}},\ and\
  \bibinfo {author} {\bibfnamefont {P.~M.}\ \bibnamefont {Zhang}},\ }\bibfield
  {title} {\bibinfo {title} {{Wigner?Souriau translations and Lorentz symmetry
  of chiral fermions}},\ }\href
  {https://doi.org/10.1016/j.physletb.2015.01.048} {\bibfield  {journal}
  {\bibinfo  {journal} {Phys. Lett. B}\ }\textbf {\bibinfo {volume} {742}},\
  \bibinfo {pages} {322} (\bibinfo {year} {2015})}\BibitemShut {NoStop}%
\bibitem [{\citenamefont {Stone}\ \emph {et~al.}(2015)\citenamefont {Stone},
  \citenamefont {Dwivedi},\ and\ \citenamefont {Zhou}}]{Stone:2015kla}%
  \BibitemOpen
  \bibfield  {author} {\bibinfo {author} {\bibfnamefont {M.}~\bibnamefont
  {Stone}}, \bibinfo {author} {\bibfnamefont {V.}~\bibnamefont {Dwivedi}},\
  and\ \bibinfo {author} {\bibfnamefont {T.}~\bibnamefont {Zhou}},\ }\bibfield
  {title} {\bibinfo {title} {{Wigner Translations and the Observer Dependence
  of the Position of Massless Spinning Particles}},\ }\href
  {https://doi.org/10.1103/PhysRevLett.114.210402} {\bibfield  {journal}
  {\bibinfo  {journal} {Phys. Rev. Lett.}\ }\textbf {\bibinfo {volume} {114}},\
  \bibinfo {pages} {210402} (\bibinfo {year} {2015})}\BibitemShut {NoStop}%
\bibitem [{\citenamefont {Chen}\ \emph {et~al.}(2015)\citenamefont {Chen},
  \citenamefont {Son},\ and\ \citenamefont {Stephanov}}]{Chen:2015gta}%
  \BibitemOpen
  \bibfield  {author} {\bibinfo {author} {\bibfnamefont {J.-Y.}\ \bibnamefont
  {Chen}}, \bibinfo {author} {\bibfnamefont {D.~T.}\ \bibnamefont {Son}},\ and\
  \bibinfo {author} {\bibfnamefont {M.~A.}\ \bibnamefont {Stephanov}},\
  }\bibfield  {title} {\bibinfo {title} {{Collisions in Chiral Kinetic
  Theory}},\ }\href {https://doi.org/10.1103/PhysRevLett.115.021601} {\bibfield
   {journal} {\bibinfo  {journal} {Phys. Rev. Lett.}\ }\textbf {\bibinfo
  {volume} {115}},\ \bibinfo {pages} {021601} (\bibinfo {year}
  {2015})}\BibitemShut {NoStop}%
\bibitem [{\citenamefont {Hidaka}\ \emph {et~al.}(2017)\citenamefont {Hidaka},
  \citenamefont {Pu},\ and\ \citenamefont {Yang}}]{Hidaka:2016yjf}%
  \BibitemOpen
  \bibfield  {author} {\bibinfo {author} {\bibfnamefont {Y.}~\bibnamefont
  {Hidaka}}, \bibinfo {author} {\bibfnamefont {S.}~\bibnamefont {Pu}},\ and\
  \bibinfo {author} {\bibfnamefont {D.-L.}\ \bibnamefont {Yang}},\ }\bibfield
  {title} {\bibinfo {title} {{Relativistic Chiral Kinetic Theory from Quantum
  Field Theories}},\ }\href {https://doi.org/10.1103/PhysRevD.95.091901}
  {\bibfield  {journal} {\bibinfo  {journal} {Phys. Rev. D}\ }\textbf {\bibinfo
  {volume} {95}},\ \bibinfo {pages} {091901} (\bibinfo {year}
  {2017})}\BibitemShut {NoStop}%
\bibitem [{\citenamefont {Vilenkin}(1979)}]{Vilenkin:1979ui}%
  \BibitemOpen
  \bibfield  {author} {\bibinfo {author} {\bibfnamefont {A.}~\bibnamefont
  {Vilenkin}},\ }\bibfield  {title} {\bibinfo {title} {{Macroscopic
  parity-violating effects: Neutrino fluxes from rotating black holes and in
  rotating thermal radiation}},\ }\href
  {https://doi.org/10.1103/PhysRevD.20.1807} {\bibfield  {journal} {\bibinfo
  {journal} {Phys. Rev. D}\ }\textbf {\bibinfo {volume} {20}},\ \bibinfo
  {pages} {1807} (\bibinfo {year} {1979})}\BibitemShut {NoStop}%
\bibitem [{\citenamefont {Son}\ and\ \citenamefont
  {Surowka}(2009)}]{Son:2009tf}%
  \BibitemOpen
  \bibfield  {author} {\bibinfo {author} {\bibfnamefont {D.~T.}\ \bibnamefont
  {Son}}\ and\ \bibinfo {author} {\bibfnamefont {P.}~\bibnamefont {Surowka}},\
  }\bibfield  {title} {\bibinfo {title} {{Hydrodynamics with Triangle
  Anomalies}},\ }\href {https://doi.org/10.1103/PhysRevLett.103.191601}
  {\bibfield  {journal} {\bibinfo  {journal} {Phys. Rev. Lett.}\ }\textbf
  {\bibinfo {volume} {103}},\ \bibinfo {pages} {191601} (\bibinfo {year}
  {2009})}\BibitemShut {NoStop}%
\bibitem [{\citenamefont {Kharzeev}\ and\ \citenamefont
  {Son}(2011)}]{Kharzeev:2010gr}%
  \BibitemOpen
  \bibfield  {author} {\bibinfo {author} {\bibfnamefont {D.~E.}\ \bibnamefont
  {Kharzeev}}\ and\ \bibinfo {author} {\bibfnamefont {D.~T.}\ \bibnamefont
  {Son}},\ }\bibfield  {title} {\bibinfo {title} {{Testing the chiral magnetic
  and chiral vortical effects in heavy ion collisions}},\ }\href
  {https://doi.org/10.1103/PhysRevLett.106.062301} {\bibfield  {journal}
  {\bibinfo  {journal} {Phys. Rev. Lett.}\ }\textbf {\bibinfo {volume} {106}},\
  \bibinfo {pages} {062301} (\bibinfo {year} {2011})}\BibitemShut {NoStop}%
\bibitem [{\citenamefont {Kharzeev}\ \emph {et~al.}(2016)\citenamefont
  {Kharzeev}, \citenamefont {Liao}, \citenamefont {Voloshin},\ and\
  \citenamefont {Wang}}]{Kharzeev:2015znc}%
  \BibitemOpen
  \bibfield  {author} {\bibinfo {author} {\bibfnamefont {D.~E.}\ \bibnamefont
  {Kharzeev}}, \bibinfo {author} {\bibfnamefont {J.}~\bibnamefont {Liao}},
  \bibinfo {author} {\bibfnamefont {S.~A.}\ \bibnamefont {Voloshin}},\ and\
  \bibinfo {author} {\bibfnamefont {G.}~\bibnamefont {Wang}},\ }\bibfield
  {title} {\bibinfo {title} {{Chiral magnetic and vortical effects in
  high-energy nuclear collisions?A status report}},\ }\href
  {https://doi.org/10.1016/j.ppnp.2016.01.001} {\bibfield  {journal} {\bibinfo
  {journal} {Prog. Part. Nucl. Phys.}\ }\textbf {\bibinfo {volume} {88}},\
  \bibinfo {pages} {1} (\bibinfo {year} {2016})}\BibitemShut {NoStop}%
\bibitem [{\citenamefont {Liu}\ \emph {et~al.}(2019)\citenamefont {Liu},
  \citenamefont {Sun},\ and\ \citenamefont {Ko}}]{Liu:2019krs}%
  \BibitemOpen
  \bibfield  {author} {\bibinfo {author} {\bibfnamefont {S.~Y.~F.}\
  \bibnamefont {Liu}}, \bibinfo {author} {\bibfnamefont {Y.}~\bibnamefont
  {Sun}},\ and\ \bibinfo {author} {\bibfnamefont {C.~M.}\ \bibnamefont {Ko}},\
  }\bibfield  {title} {\bibinfo {title} {{Spin polarizations in a covariant
  angular momentum conserved chiral transport model}},\ }\href@noop {} {\
  (\bibinfo {year} {2019})},\ \Eprint {https://arxiv.org/abs/1910.06774}
  {arXiv:1910.06774 [nucl-th]} \BibitemShut {NoStop}%
\bibitem [{\citenamefont {Becattini}\ and\ \citenamefont
  {Karpenko}(2018)}]{Becattini:2017gcx}%
  \BibitemOpen
  \bibfield  {author} {\bibinfo {author} {\bibfnamefont {F.}~\bibnamefont
  {Becattini}}\ and\ \bibinfo {author} {\bibfnamefont {I.}~\bibnamefont
  {Karpenko}},\ }\bibfield  {title} {\bibinfo {title} {{Collective Longitudinal
  Polarization in Relativistic Heavy-Ion Collisions at Very High Energy}},\
  }\href {https://doi.org/10.1103/PhysRevLett.120.012302} {\bibfield  {journal}
  {\bibinfo  {journal} {Phys. Rev. Lett.}\ }\textbf {\bibinfo {volume} {120}},\
  \bibinfo {pages} {012302} (\bibinfo {year} {2018})}\BibitemShut {NoStop}%
\bibitem [{\citenamefont {Niida}(2019)}]{Niida:2018hfw}%
  \BibitemOpen
  \bibfield  {author} {\bibinfo {author} {\bibfnamefont {T.}~\bibnamefont
  {Niida}} (\bibinfo {collaboration} {STAR}),\ }\bibfield  {title} {\bibinfo
  {title} {{Global and local polarization of $\Lambda$ hyperons in Au+Au
  collisions at 200 GeV from STAR}},\ }\href
  {https://doi.org/10.1016/j.nuclphysa.2018.08.034} {\bibfield  {journal}
  {\bibinfo  {journal} {Nucl. Phys. A}\ }\textbf {\bibinfo {volume} {982}},\
  \bibinfo {pages} {511} (\bibinfo {year} {2019})}\BibitemShut {NoStop}%
\bibitem [{\citenamefont {Adam}\ \emph {et~al.}(2019)\citenamefont {Adam} \emph
  {et~al.}}]{Adam:2019srw}%
  \BibitemOpen
  \bibfield  {author} {\bibinfo {author} {\bibfnamefont {J.}~\bibnamefont
  {Adam}} \emph {et~al.} (\bibinfo {collaboration} {STAR}),\ }\bibfield
  {title} {\bibinfo {title} {{Polarization of $\Lambda$ ($\bar{\Lambda}$)
  hyperons along the beam direction in Au+Au collisions at $\sqrt{s_{_{NN}}}$ =
  200 GeV}},\ }\href {https://doi.org/10.1103/PhysRevLett.123.132301}
  {\bibfield  {journal} {\bibinfo  {journal} {Phys. Rev. Lett.}\ }\textbf
  {\bibinfo {volume} {123}},\ \bibinfo {pages} {132301} (\bibinfo {year}
  {2019})}\BibitemShut {NoStop}%
\bibitem [{\citenamefont {Born}\ and\ \citenamefont
  {Infeld}(1935)}]{born1935quantization}%
  \BibitemOpen
  \bibfield  {author} {\bibinfo {author} {\bibfnamefont {M.}~\bibnamefont
  {Born}}\ and\ \bibinfo {author} {\bibfnamefont {L.}~\bibnamefont {Infeld}},\
  }\bibfield  {title} {\bibinfo {title} {On the quantization of the new field
  theory ii},\ }\href@noop {} {\bibfield  {journal} {\bibinfo  {journal}
  {Proceedings of the Royal Society of London. Series A-Mathematical and
  Physical Sciences}\ }\textbf {\bibinfo {volume} {150}},\ \bibinfo {pages}
  {141} (\bibinfo {year} {1935})}\BibitemShut {NoStop}%
\bibitem [{\citenamefont {Pryce}(1948)}]{pryce1948mass}%
  \BibitemOpen
  \bibfield  {author} {\bibinfo {author} {\bibfnamefont {M.~H.~L.}\
  \bibnamefont {Pryce}},\ }\bibfield  {title} {\bibinfo {title} {The
  mass-centre in the restricted theory of relativity and its connexion with the
  quantum theory of elementary particles},\ }\href@noop {} {\bibfield
  {journal} {\bibinfo  {journal} {Proceedings of the Royal Society of London.
  Series A. Mathematical and Physical Sciences}\ }\textbf {\bibinfo {volume}
  {195}},\ \bibinfo {pages} {62} (\bibinfo {year} {1948})}\BibitemShut
  {NoStop}%
\bibitem [{\citenamefont {M{\o}ller}(1949)}]{moller1949definition}%
  \BibitemOpen
  \bibfield  {author} {\bibinfo {author} {\bibfnamefont {C.}~\bibnamefont
  {M{\o}ller}},\ }\bibfield  {title} {\bibinfo {title} {On the definition of
  the centre of gravity of an arbitrary closed system in the theory of
  relativity},\ }\href@noop {} {\bibfield  {journal} {\bibinfo  {journal}
  {Communications of the Dublin Institute for Advanced Studies}\ } (\bibinfo
  {year} {1949})}\BibitemShut {NoStop}%
\bibitem [{\citenamefont {Newton}\ and\ \citenamefont
  {Wigner}(1949)}]{Newton:1949cq}%
  \BibitemOpen
  \bibfield  {author} {\bibinfo {author} {\bibfnamefont {T.~D.}\ \bibnamefont
  {Newton}}\ and\ \bibinfo {author} {\bibfnamefont {E.~P.}\ \bibnamefont
  {Wigner}},\ }\bibfield  {title} {\bibinfo {title} {{Localized States for
  Elementary Systems}},\ }\href {https://doi.org/10.1103/RevModPhys.21.400}
  {\bibfield  {journal} {\bibinfo  {journal} {Rev. Mod. Phys.}\ }\textbf
  {\bibinfo {volume} {21}},\ \bibinfo {pages} {400} (\bibinfo {year}
  {1949})}\BibitemShut {NoStop}%
\bibitem [{\citenamefont {Fleming}(1965)}]{PhysRev.137.B188}%
  \BibitemOpen
  \bibfield  {author} {\bibinfo {author} {\bibfnamefont {G.~N.}\ \bibnamefont
  {Fleming}},\ }\bibfield  {title} {\bibinfo {title} {Covariant position
  operators, spin, and locality},\ }\href
  {https://doi.org/10.1103/PhysRev.137.B188} {\bibfield  {journal} {\bibinfo
  {journal} {Phys. Rev.}\ }\textbf {\bibinfo {volume} {137}},\ \bibinfo {pages}
  {B188} (\bibinfo {year} {1965})}\BibitemShut {NoStop}%
\bibitem [{\citenamefont {Bliokh}\ \emph {et~al.}(2017)\citenamefont {Bliokh},
  \citenamefont {Dennis},\ and\ \citenamefont {Nori}}]{bliokh2017position}%
  \BibitemOpen
  \bibfield  {author} {\bibinfo {author} {\bibfnamefont {K.~Y.}\ \bibnamefont
  {Bliokh}}, \bibinfo {author} {\bibfnamefont {M.~R.}\ \bibnamefont {Dennis}},\
  and\ \bibinfo {author} {\bibfnamefont {F.}~\bibnamefont {Nori}},\ }\bibfield
  {title} {\bibinfo {title} {Position, spin, and orbital angular momentum of a
  relativistic electron},\ }\href@noop {} {\bibfield  {journal} {\bibinfo
  {journal} {Physical Review A}\ }\textbf {\bibinfo {volume} {96}},\ \bibinfo
  {pages} {023622} (\bibinfo {year} {2017})}\BibitemShut {NoStop}%
\bibitem [{\citenamefont {Lorcé}(2018)}]{Lorce:2018zpf}%
  \BibitemOpen
  \bibfield  {author} {\bibinfo {author} {\bibfnamefont {C.}~\bibnamefont
  {Lorcé}},\ }\bibfield  {title} {\bibinfo {title} {{The relativistic center
  of mass in field theory with spin}},\ }\href
  {https://doi.org/10.1140/epjc/s10052-018-6249-3} {\bibfield  {journal}
  {\bibinfo  {journal} {Eur. Phys. J. C}\ }\textbf {\bibinfo {volume} {78}},\
  \bibinfo {pages} {785} (\bibinfo {year} {2018})},\ \Eprint
  {https://arxiv.org/abs/1805.05284} {arXiv:1805.05284 [hep-ph]} \BibitemShut
  {NoStop}%
\bibitem [{\citenamefont {Hattori}\ \emph {et~al.}(2019)\citenamefont
  {Hattori}, \citenamefont {Hidaka},\ and\ \citenamefont
  {Yang}}]{Hattori:2019ahi}%
  \BibitemOpen
  \bibfield  {author} {\bibinfo {author} {\bibfnamefont {K.}~\bibnamefont
  {Hattori}}, \bibinfo {author} {\bibfnamefont {Y.}~\bibnamefont {Hidaka}},\
  and\ \bibinfo {author} {\bibfnamefont {D.-L.}\ \bibnamefont {Yang}},\
  }\bibfield  {title} {\bibinfo {title} {{Axial Kinetic Theory and Spin
  Transport for Fermions with Arbitrary Mass}},\ }\href
  {https://doi.org/10.1103/PhysRevD.100.096011} {\bibfield  {journal} {\bibinfo
   {journal} {Phys. Rev. D}\ }\textbf {\bibinfo {volume} {100}},\ \bibinfo
  {pages} {096011} (\bibinfo {year} {2019})},\ \Eprint
  {https://arxiv.org/abs/1903.01653} {arXiv:1903.01653 [hep-ph]} \BibitemShut
  {NoStop}%
\bibitem [{\citenamefont {Peskin}\ and\ \citenamefont
  {Schroeder}(1995)}]{Peskin:1995ev}%
  \BibitemOpen
  \bibfield  {author} {\bibinfo {author} {\bibfnamefont {M.~E.}\ \bibnamefont
  {Peskin}}\ and\ \bibinfo {author} {\bibfnamefont {D.~V.}\ \bibnamefont
  {Schroeder}},\ }\href {http://www.slac.stanford.edu/~mpeskin/QFT.html} {\emph
  {\bibinfo {title} {{An Introduction to quantum field theory}}}}\ (\bibinfo
  {publisher} {Addison-Wesley},\ \bibinfo {address} {Reading, USA},\ \bibinfo
  {year} {1995})\BibitemShut {NoStop}%
\bibitem [{\citenamefont {{Luba{\'n}ski}}(1942)}]{Lubanski:1942A}%
  \BibitemOpen
  \bibfield  {author} {\bibinfo {author} {\bibfnamefont {J.~K.}\ \bibnamefont
  {{Luba{\'n}ski}}},\ }\bibfield  {title} {\bibinfo {title} {{Sur la theorie
  des particules {\'e}l{\'e}mentaires de spin quelconque. I}},\ }\href
  {https://doi.org/10.1016/S0031-8914(42)90113-7} {\bibfield  {journal}
  {\bibinfo  {journal} {Physica}\ }\textbf {\bibinfo {volume} {9}},\ \bibinfo
  {pages} {310} (\bibinfo {year} {1942})}\BibitemShut {NoStop}%
\bibitem [{\citenamefont {{Lubanski}}(1942)}]{Lubanski:1942B}%
  \BibitemOpen
  \bibfield  {author} {\bibinfo {author} {\bibfnamefont {J.~K.}\ \bibnamefont
  {{Lubanski}}},\ }\bibfield  {title} {\bibinfo {title} {{Sur la th{\'e}orie
  des particules {\'e}l{\'e}mentaires de spin quelconque. II}},\ }\href
  {https://doi.org/10.1016/S0031-8914(42)90114-9} {\bibfield  {journal}
  {\bibinfo  {journal} {Physica}\ }\textbf {\bibinfo {volume} {9}},\ \bibinfo
  {pages} {325} (\bibinfo {year} {1942})}\BibitemShut {NoStop}%
\bibitem [{\citenamefont {Weinberg}(2005)}]{Weinberg:1995mt}%
  \BibitemOpen
  \bibfield  {author} {\bibinfo {author} {\bibfnamefont {S.}~\bibnamefont
  {Weinberg}},\ }\href@noop {} {\emph {\bibinfo {title} {{The Quantum theory of
  fields. Vol. 1: Foundations}}}}\ (\bibinfo  {publisher} {Cambridge University
  Press},\ \bibinfo {year} {2005})\BibitemShut {NoStop}%
\bibitem [{\citenamefont {Florkowski}\ \emph {et~al.}(2018)\citenamefont
  {Florkowski}, \citenamefont {Friman}, \citenamefont {Jaiswal},\ and\
  \citenamefont {Speranza}}]{Florkowski:2017ruc}%
  \BibitemOpen
  \bibfield  {author} {\bibinfo {author} {\bibfnamefont {W.}~\bibnamefont
  {Florkowski}}, \bibinfo {author} {\bibfnamefont {B.}~\bibnamefont {Friman}},
  \bibinfo {author} {\bibfnamefont {A.}~\bibnamefont {Jaiswal}},\ and\ \bibinfo
  {author} {\bibfnamefont {E.}~\bibnamefont {Speranza}},\ }\bibfield  {title}
  {\bibinfo {title} {{Relativistic fluid dynamics with spin}},\ }\href
  {https://doi.org/10.1103/PhysRevC.97.041901} {\bibfield  {journal} {\bibinfo
  {journal} {Phys. Rev. C}\ }\textbf {\bibinfo {volume} {97}},\ \bibinfo
  {pages} {041901} (\bibinfo {year} {2018})}\BibitemShut {NoStop}%
\bibitem [{\citenamefont {Becattini}\ \emph {et~al.}(2019)\citenamefont
  {Becattini}, \citenamefont {Florkowski},\ and\ \citenamefont
  {Speranza}}]{Becattini:2018duy}%
  \BibitemOpen
  \bibfield  {author} {\bibinfo {author} {\bibfnamefont {F.}~\bibnamefont
  {Becattini}}, \bibinfo {author} {\bibfnamefont {W.}~\bibnamefont
  {Florkowski}},\ and\ \bibinfo {author} {\bibfnamefont {E.}~\bibnamefont
  {Speranza}},\ }\bibfield  {title} {\bibinfo {title} {{Spin tensor and its
  role in non-equilibrium thermodynamics}},\ }\href
  {https://doi.org/10.1016/j.physletb.2018.12.016} {\bibfield  {journal}
  {\bibinfo  {journal} {Phys. Lett. B}\ }\textbf {\bibinfo {volume} {789}},\
  \bibinfo {pages} {419} (\bibinfo {year} {2019})}\BibitemShut {NoStop}%
\bibitem [{\citenamefont {Weickgenannt}\ \emph {et~al.}(2019)\citenamefont
  {Weickgenannt}, \citenamefont {Sheng}, \citenamefont {Speranza},
  \citenamefont {Wang},\ and\ \citenamefont {Rischke}}]{Weickgenannt:2019dks}%
  \BibitemOpen
  \bibfield  {author} {\bibinfo {author} {\bibfnamefont {N.}~\bibnamefont
  {Weickgenannt}}, \bibinfo {author} {\bibfnamefont {X.-L.}\ \bibnamefont
  {Sheng}}, \bibinfo {author} {\bibfnamefont {E.}~\bibnamefont {Speranza}},
  \bibinfo {author} {\bibfnamefont {Q.}~\bibnamefont {Wang}},\ and\ \bibinfo
  {author} {\bibfnamefont {D.~H.}\ \bibnamefont {Rischke}},\ }\bibfield
  {title} {\bibinfo {title} {{Kinetic theory for massive spin-1/2 particles
  from the Wigner-function formalism}},\ }\href
  {https://doi.org/10.1103/PhysRevD.100.056018} {\bibfield  {journal} {\bibinfo
   {journal} {Phys. Rev. D}\ }\textbf {\bibinfo {volume} {100}},\ \bibinfo
  {pages} {056018} (\bibinfo {year} {2019})}\BibitemShut {NoStop}%
\bibitem [{\citenamefont {Gao}\ and\ \citenamefont
  {Liang}(2019)}]{Gao:2019znl}%
  \BibitemOpen
  \bibfield  {author} {\bibinfo {author} {\bibfnamefont {J.-H.}\ \bibnamefont
  {Gao}}\ and\ \bibinfo {author} {\bibfnamefont {Z.-T.}\ \bibnamefont
  {Liang}},\ }\bibfield  {title} {\bibinfo {title} {{Relativistic Quantum
  Kinetic Theory for Massive Fermions and Spin Effects}},\ }\href@noop {}
  {\bibfield  {journal} {\bibinfo  {journal} {Phys. Rev. D}\ }\textbf {\bibinfo
  {volume} {100}},\ \bibinfo {pages} {056021} (\bibinfo {year}
  {2019})}\BibitemShut {NoStop}%
\bibitem [{\citenamefont {Yang}\ \emph {et~al.}(2020)\citenamefont {Yang},
  \citenamefont {Hattori},\ and\ \citenamefont {Hidaka}}]{Yang:2020hri}%
  \BibitemOpen
  \bibfield  {author} {\bibinfo {author} {\bibfnamefont {D.-L.}\ \bibnamefont
  {Yang}}, \bibinfo {author} {\bibfnamefont {K.}~\bibnamefont {Hattori}},\ and\
  \bibinfo {author} {\bibfnamefont {Y.}~\bibnamefont {Hidaka}},\ }\bibfield
  {title} {\bibinfo {title} {{Quantum kinetic theory for spin transport:
  general formalism for collisional effects}},\ }\href@noop {} {\  (\bibinfo
  {year} {2020})},\ \Eprint {https://arxiv.org/abs/2002.02612}
  {arXiv:2002.02612 [hep-ph]} \BibitemShut {NoStop}%
\bibitem [{\citenamefont {Bhadury}\ \emph {et~al.}(2020)\citenamefont
  {Bhadury}, \citenamefont {Florkowski}, \citenamefont {Jaiswal}, \citenamefont
  {Kumar},\ and\ \citenamefont {Ryblewski}}]{Bhadury:2020puc}%
  \BibitemOpen
  \bibfield  {author} {\bibinfo {author} {\bibfnamefont {S.}~\bibnamefont
  {Bhadury}}, \bibinfo {author} {\bibfnamefont {W.}~\bibnamefont {Florkowski}},
  \bibinfo {author} {\bibfnamefont {A.}~\bibnamefont {Jaiswal}}, \bibinfo
  {author} {\bibfnamefont {A.}~\bibnamefont {Kumar}},\ and\ \bibinfo {author}
  {\bibfnamefont {R.}~\bibnamefont {Ryblewski}},\ }\bibfield  {title} {\bibinfo
  {title} {{Relativistic dissipative spin dynamics in the relaxation time
  approximation}},\ }\href@noop {} {\  (\bibinfo {year} {2020})},\ \Eprint
  {https://arxiv.org/abs/2002.03937} {arXiv:2002.03937 [hep-ph]} \BibitemShut
  {NoStop}%
\end{thebibliography}%
\end{document}